\documentclass[preprint]{elsarticle}
\usepackage[utf8]{inputenc}

\usepackage{geometry}
\usepackage{amssymb}
\usepackage{appendix}
\usepackage{amsmath}
\usepackage{bm}
\usepackage{float}
\usepackage{graphicx}

\usepackage{siunitx}
\usepackage{natbib}
\setcitestyle{authoryear,round,semicolon}

\usepackage{caption}
\usepackage{subcaption}
\usepackage{hyperref}
\hypersetup{colorlinks = true, linkcolor = blue, filecolor = magenta, urlcolor = cyan}

\newcommand{\Rey}{\mathrm{Re}}
\newcommand{\Ca}{\mathrm{Ca}}

\begin{document}

\begin{frontmatter}

\title{Characterizing viscoelastic materials via ensemble-based data assimilation of bubble collapse observations}

\author[add1]{Jean-Sebastien~Spratt\corref{cor}}
\cortext[cor]{Corresponding author; jspratt@caltech.edu}
\author[add1]{Mauro~Rodriguez}
\author[add1]{Kevin~Schmidmayer}
\author[add1]{Spencer H.~Bryngelson}
\author[add2]{Jin~Yang}
\author[add2]{Christian~Franck}
\author[add1]{Tim~Colonius}

\address[add1]{Division of Engineering and Applied Science, California Institute of Technology, Pasadena, CA 91125, USA}
\address[add2]{Department of Mechanical Engineering, University of Wisconsin-Madison, Madison, WI 53706, USA}

\begin{abstract}
    Viscoelastic material properties at high strain rates are needed to model many biological and medical systems.
    Bubble cavitation can induce such strain rates, and the resulting bubble dynamics are sensitive to the material properties.
    Thus, in principle, these properties can be inferred via measurements of the bubble dynamics.
    \citet{Estrada_2018} demonstrated such bubble-dynamic high-strain-rate rheometry by using least-squares shooting to minimize the difference between simulated and experimental bubble radius histories.
    We generalize their technique to account for additional uncertainties in the model, initial conditions, and material properties needed to uniquely simulate the bubble dynamics.
    Ensemble-based data assimilation minimizes the computational expense associated with the bubble cavitation model.
    We test an ensemble Kalman filter (EnKF), an iterative ensemble Kalman smoother (IEnKS), and a hybrid ensemble-based 4D--Var method (En4D--Var) on synthetic data, assessing their estimations of the viscosity and shear modulus of a Kelvin--Voigt material.
    Results show that En4D--Var and IEnKS provide better moduli estimates than EnKF.
    Applying these methods to the experimental data of \citet{Estrada_2018} yields similar material property estimates to those they obtained, but provides additional information about uncertainties.
    In particular, the En4D--Var yields lower viscosity estimates for some experiments, and the dynamic estimators reveal a potential mechanism that is unaccounted for in the model, whereby the viscosity is reduced in some cases due to material damage occurring at bubble collapse.
\end{abstract}


\journal{Journal of the Mechanics and Physics of Solids}

\end{frontmatter}

%%%%%%%%%%%%%%%%%%%%%%%%%%%%%%
%%%%%%%% INTRODUCTION %%%%%%%%
%%%%%%%%%%%%%%%%%%%%%%%%%%%%%%

\section{Introduction}

Measuring the mechanical properties of soft viscoelastic materials at high strain rates (exceeding $\SI{e3}{\per\second})$ is a challenging goal of rheometry.
These measurements are of particular interest in biological and medical engineering, where high strain rates occur during impact and blast exposure~\citep{Bar-Kochba_2016, Sartinoranont_2012, Meaney_2011, Nyein_2010} and during therapeutic ultrasound~\citep{Maxwell_2009, Xu_2007, Mancia_2017, Vlaisavljevich_2015, Bailey_2003}.
Cavitation, which can take place on exposure to tensile waves, induces high strain rates in surrounding materials, and the resulting bubble dynamics are sensitive to the adjacent material properties \citep{Estrada_2018}.
This observation led \citet{Estrada_2018} to propose a high-strain rate rheometer to estimate the viscoelastic properties of polyacrylamide gels through observation of the bubble radius time history during a laser-generated cavitation event in a sample of the material.
To infer the viscosity and shear modulus, they developed a least-square fitting technique which minimizes the difference between the measurements and the bubble radius history predicted through a mechanical model of a spherical bubble in an assumed neo-Hookean Kelvin--Voigt viscoelastic material.

In this paper, we generalize bubble-dynamic rheometry by considering data assimilation (DA) techniques that can potentially improve predictions in uncertainty-prone high-strain-rate regimes.
Provided the material properties are observable and the dynamics are sufficiently sensitive to their values, DA provides a solution to this inverse problem that accounts for uncertainty in both the model and data~\citep{Evensen_2009, Evensen_2009_2, Bocquet_2013_2, Schillings_2017}.
Thus, for bubble-dynamics-based rheometry, DA can address additional uncertainties beyond the unknown viscosity and shear modulus, including those associated with measurement noise, additional material properties, modeling assumptions, and initial conditions.
DA techniques are characterized as {\it filters} if the state (and parameters) are updated at each moment based on the prior trajectory or {\it smoothers} if the state history and parameters are estimated over a horizon.
By considering both filters and smoothers, we can gain additional insights into whether constant or time-varying parameters best fit the observed behavior.

Three data assimilation methods are considered: an ensemble Kalman filter (EnKF), an iterative ensemble Kalman smoother (IEnKS), and a hybrid ensemble-based 4D--Var method (En4D--Var).
EnKF and IEnKS are variations on the classical Kalman Filter (KF)~\citep{Kalman_1960}.
Their algorithms follow the same structure as the KF, which assimilates data at each step of a discrete time series.
These are Monte Carlo methods that represent the system via an ensemble.
The assimilation uses statistics of the ensemble to calculate a sample covariance.
This replaces the covariance matrix of the KF and thus the covariance forecast operator, reducing computational cost.
En4D--Var is a different ensemble-based approach to the assimilation problem.
It is a variant of offline 4D variational data assimilation (4D--Var)~\citep{Caya_2005, Tremolet_2007}, where a guess for the initial condition is iterated upon to improve the fit to the data over the entire time domain.

In section~\ref{sec:physical_model}, we describe the specific material--bubble-dynamic model, which matches the one considered by \citet{Estrada_2018}.
In section~\ref{sec:da}, details of the data assimilation algorithms and their implementation are provided.
We then examine the relative merits of the estimators in section~\ref{sec:results_sim} using synthetic data generated by running the model (with additional simulated noise).
This allows us to gauge their relative performance for cases with no modeling uncertainties.
Next, in section~\ref{sec:exp}, we apply these estimators to established experimental data for polyacrylamide gels.
We compare and contrast results with each method and with the the estimates of~\citet{Estrada_2018}.
We summarize the conclusions in section~\ref{sec:conclusion}.

%%%%%%%%%%%%%%%%%%%%%%%%%%%%%%
 %%% Bubble Dynamics Model %%%
%%%%%%%%%%%%%%%%%%%%%%%%%%%%%%

\section{Bubble dynamics model} \label{sec:physical_model}

A physical model for the collapsing bubbles is required to characterize the viscoelastic properties of surrounding materials.
Many spherical bubble dynamics models exist.
Of particular relevance here are those for cavitation in soft materials~\citep{Gaudron_2015, Yang_2005} and specific numerical methods for solving them~\citep{Warnez_2015, Barajas_2017}.
We use the model of \citet{Estrada_2018}, which adopts approximations validated in previous spherical-bubble models~\citep{Prosperetti_1986, Prosperetti_1988, Akhatov_2001, Epstein_1972, Keller_1980, Preston_2007}.
Key assumptions of this model are that the motion of the bubble and its contents are spherically symmetric, the bubble pressure is spatially uniform (homobaricity), the temperature of the surrounding material is constant, and that there is no mass transfer of the non-condensible gas across the bubble wall.

The Keller--Miksis equation models the radius evolution~\citep{Keller_1980},
\begin{gather} 
    \left(1-\frac{\dot{R}}{c}\right)R\ddot{R} + 
    \frac{3}{2}\left(1-\frac{\dot{R}}{3c}\right)\dot{R}^2 = 
    \frac{1}{\rho}\left(1+\frac{\dot{R}}{c}\right)\left(p_b - \frac{2\gamma}{R} + 
    S - p_{\infty}\right) + 
    \frac{1}{\rho}\frac{R}{c}
    \dot{\overline{\left(p_b - \frac{2\gamma}{R} + S \right)}},
    \label{eq:R}
\end{gather}
where $R$ is the bubble radius, $c$ the material speed of sound, $\rho$ the material density, $p_b$ the bubble internal pressure, $\gamma$ the bubble-wall surface tension, $S$ the stress integral (see~\eqref{eq:stress_integral}), and $p_\infty$ the far-field pressure.
Under the model assumptions, no mass or energy conservation equations are needed outside the bubble.
Furthermore, the conservation of momentum simplifies to an ordinary differential equation for the bubble pressure 
\begin{gather} \label{eq:pres}
    \dot{p}_b = \frac{3}{R}\left[-\kappa p_b \dot{R} + (\kappa - 1)K(T(R)) \left.
    \frac{\partial T}{\partial r}\right|_{r=R} + 
    \kappa p_b \frac{C_{p,v}}{C_p(k_v(R))}\frac{D}{1-k_v(R)}\left. 
    \frac{\partial k_v}{\partial r}\right|_{r=R}\right],
\end{gather}
where $\kappa$ is the specific heat ratio, $K$ the thermal conductivity, $T$ the gas temperature, $C_p$ the specific heat, $D$ the binary diffusion coefficient, and $k_v$ the vapor mass fraction.
Subscripts $g$, $v$, and $m$ refer to gas, vapor, and mixture properties.
Conservation of energy in the bubble interior yields an equation for the bubble temperature:
\begin{gather} \label{eq:temp}
    \rho_mC_p\left(\frac{\partial T}{\partial t} + 
    v_m\frac{\partial T}{\partial r} \right) = 
    \dot{p}_b + 
    \frac{1}{r^2}\frac{\partial}{\partial r}
    \left(r^2K\frac{\partial T}{\partial r}\right) + 
    \rho_m (C_{p,v}-C_{p,g}) 
    D \frac{\partial k_v}{\partial r}\frac{\partial T}{\partial r},
\end{gather}
where $v_m$ is the radial mixture velocity and $\rho_m$ the mixture density.
The boundary condition $T(R) = T_{\infty}$ follows from the model assumptions.
The radial mixture velocities are computed as
\begin{gather}
    v_m(r,t)=\frac{1}{\kappa p_b}\left[(\kappa -1)K\frac{\partial T}{\partial R} - \frac{1}{3}r\dot{p}_b\right]+\frac{C_{p,v} - C_{p,g}}{C_{p,m}}D\frac{\partial k}{\partial r}.
\end{gather}
Fick's law describes the mass diffusion process in the bubble.
Casting the conservation of mass inside the bubble in terms of the mixture density, the vapor mass fraction inside the bubble is
\begin{gather} \label{eq:mass}
    \frac{\partial k_v}{\partial t} + v_m \frac{\partial k_v}{\partial r} = 
    \frac{1}{\rho_m}\frac{1}{r^2}
    \frac{\partial}{\partial r}
    \left(r^2\rho_m D\frac{\partial k_v}{\partial r}\right).
\end{gather}
Under the assumption of equilibrium phase change at the bubble wall, the associated boundary condition at the wall is $p_{v,sat}(T(R)) = \mathcal{R}_v k(R) \rho_m(R)T(R)$, where $p_{v,sat}$ is the saturation pressure of the vapor and $\mathcal{R}_v$ is the gas constant of the vapor.

Equations \eqref{eq:R}, \eqref{eq:pres}, \eqref{eq:temp}, and \eqref{eq:mass} form a system of equations.
This system is evolved in time with an implicit Runge--Kutta algorithm that uses the trapezoidal rule and backwards differentiation at each step (TR--BDF2)~\citep{Hosea_1996}.
The partial differential equations for temperature and vapor mass fraction are discretized in space via a uniform grid and computed using second-order-accurate central finite differences.
\citet{Estrada_2018} showed that the finite-deformation neo-Hookean Kelvin--Voigt model can represent the material response at high strain rates.
In this framework, the material is modeled with a parallel spring (neo-Hookean elastic response with shear modulus $G$) and dashpot (linear viscous response with viscosity $\mu$).
The stress integral in~\eqref{eq:R} is
\begin{gather} \label{eq:stress_integral}
    S = -\frac{G}{2}\left[5-\left(\frac{R_0}{R}\right)^4-4\frac{R_0}{R}\right]-\frac{4\mu \dot{R}}{R},
\end{gather}
where $R_0$ is the equilibrium bubble radius~\cite{Gaudron_2015}.

%%%%%%%%%%%%%%%%%%%%%%%%%%%%%%
% Data Assimilation Methods %%
%%%%%%%%%%%%%%%%%%%%%%%%%%%%%%

\section{Data assimilation methods} \label{sec:da}

Two difficulties that drive the choice of data assimilation method are the nonlinearity of the dynamics and large state vector required to discretize the partial differential equations adequately.
The former rules out the standard linearized Kalman filter (EKF)~\citep{Kalman_1960} and the latter renders its direct nonlinear extensions (e.g.\ the unscented Kalman filter, UKF) computationally prohibitive.
Instead, ensemble-based methods~\citep{Evensen_1994} are considered.
They combine computational efficiency with nonlinear dynamics by approximating the state covariance via statistics of a finite (and typically small) ensemble.
We consider three specific ensemble methods: an ensemble Kalman filter (EnKF), an iterative ensemble Kalman smoother (IEnKS) and a hybrid ensemble-based 4D--Var method.

The discretized equations of section~\ref{sec:physical_model} are re-written as a nonlinear operator $F$, and we define the linear observation function $H$ that maps the state $\bm{x}$ to measurement space.
This yields the discrete-time dynamical system
\begin{align}
    \bm{x}_{k+1} &= F(\bm{x}_k)+ \bm{\eta}_k \label{eq:forecast model}, \\
    \bm{y}_k &= H(\bm{x}_k) + \bm{\nu}_k \label{eq:observation model},
\end{align}
where
\begin{align*}
    \bm{x}_k &\in \mathbb{R}^d \ , \ \bm{y}_k \in \mathbb{R}^n, \\
    \bm{\eta}_k &\sim \mathcal{N}(0,\Sigma) \ , \ \bm{\nu}_k \sim \mathcal{N}(0,\Gamma), \\
    F&:\mathbb{R}^d \rightarrow \mathbb{R}^d \ , \ H:\mathbb{R}^d \rightarrow \mathbb{R}^n.
\end{align*}
$\bm{x}_k$ is the $d$-dimensional state comprised of all the dependent variables plus the unknown parameters
\begin{gather}
    \bm{x} = \{ R, \dot{R}, p_b, 
    S, \mathbf{T}, \mathbf{C}, 
    \log(\Ca), \log(\Rey) \},
\end{gather}
which are the bubble-wall radius, velocity, bubble pressure, stress integral, the discretized temperature and vapor concentration fields inside the bubble, and the log-Cauchy and log-Reynolds numbers, respectively.
The Cauchy and Reynolds numbers are defined as
\begin{gather}
    \Ca \equiv \frac{p_\infty}{G} 
    \quad \text{and} \quad
    \Rey \equiv \frac{\sqrt{\rho p_\infty}R_{\mathrm{max}}}{\mu} \label{eq:ReCa}.
\end{gather}
These quantities appear in the nondimensionalized model equations of section~\ref{sec:physical_model} and the shear modulus $G$ and viscosity $\mu$ can be computed via~\eqref{eq:ReCa}.
The forecast operator $F$  maps $\log(\Ca)$ and $\log(\Rey)$ to themselves because they are constant in the physical model.
Using the logarithm avoids negative (and thus non-physical) values during the analysis step of the assimilation algorithms (described in sections~\ref{sec:EnKF}, \ref{sec:IEnKS}, and \ref{sec:En4D-Var}).

The variable $\bm{y}_k$ is the $n$-dimensional observation (data) at time $k$.
$\bm{\eta}_k$ is the unknown process noise (or model error) added to $H(\bm{x}_k)$ to retrieve $\bm{y}_k$.
It is assumed to be Gaussian with zero mean and standard deviation $\Sigma$.
Similarly, $\bm{\nu}_k$ is the assumed Gaussian measurement noise added to $F(\bm{x}_k)$ to obtain $\bm{x}_{k+1}$, with zero mean and unknown standard deviation $\Gamma$.
Throughout this study, the only available measurement is the bubble radius.
This means that $\bm{y}_k$ is the radius only, and the observation operator $H$ is the linear map from the state vector to its first element $R$.
In the following, the linear operator $H$ is sometimes represented as the matrix $\bm{H}$ for clarity ($\bm{Hx}=H(\bm{x})$).

The following methods estimate the full state vector $\bm{x}$ (including parameters of interest $\log(\Ca)$ and $\log(\Rey)$) based on observations of $\bm{y}$.
The EnKF and IEnKS are online (or quasi-online) methods---they optimize the value of $\bm{x}$ at each time through the simulation.
The IEnKS is deemed quasi-online because it uses data from future times as well.
The estimation trails the simulation time by a fixed number of time steps called the lag.
Alternatively, the En4D--Var is an offline method, which only optimizes the initial condition for $\bm{x}$, taking into account data from the entire time-domain.

%%%%

\subsection{The ensemble Kalman filter} \label{sec:EnKF}

The ensemble Kalman filter~\citep{Evensen_1994} represents the probability density function (PDF) for the state of the dynamics through the statistics of an ensemble of $q$ state vectors.
It does not require an adjoint, or deriving a tangent linear operator to the physical model~\citep{Evensen_2003,Evensen_2009}.
Starting with suitably randomized initial conditions, each ensemble member is propagated through the physical model, and the predictions are then corrected using the ensemble statistics.
The ensemble is initialized with a guess for the initial condition $\bm{x}_0$ as the mean, and a given covariance corresponding to the expected error covariance.
In practice, each ensemble member is independently sampled from a normal distribution with mean $\bm{x}_0$ and the assumed covariance matrix.
Several initialization strategies exist depending on the system and its dynamics.
In the present case, the nonlinear dynamics render a systematic approach difficult.
Instead, the ensemble is initialized with a covariance corresponding to our best estimate based on simulations, and adjusted through trial and error to optimize results.
An ensemble size of $q=48$ is used for the tests.
This has shown to give accurate results while keeping computational costs modest.
At any given time, the estimated value for the state vector is then taken to be the ensemble average.
\begin{gather} \label{eq:x_bar}
    \overline{\bm{x}}_k = \frac{1}{q}\sum^q_{j=1}\bm{x}_k^{(j)}.
\end{gather}

The filter is broken down into a forecast and an analysis step.
In the forecast step, the physical model is used to step the state forward in time with~\eqref{eq:forecast model}.
Each representation of the state vector $\bm{x}_k^{(j)}$ in the ensemble at time $k$ is propagated through $F$ with $\hat{\bm{x}}_{k+1}^{(j)} = F(\bm{x}_k^{(j)})$.
Next, in the analysis step, if an experimental measurement $\bm{y}_{k+1}$ is available at the current time step $k+1$, then it is used to correct the forecast.
As described in the dynamical system equations, each ensemble member is mapped to measurement space $H(\bm{x}_{k+1})$.
The analysis proceeds by minimizing a cost function involving the difference between $H(\bm{x})$ and the data point $\bm{y}$, while accounting for measurement noise and model error.
This cost function marks the key difference between the EnKF and other ensemble Kalman methods.
The EnKF cost function is given by
\begin{align}
    J(\bm{x}) &= \frac{1}{2} \lVert \bm{y}_k-H(\bm{x})\rVert^2_{\bm{R}} + \frac{1}{2}\lVert \bm{x}-\hat{\bm{x}_k}\rVert^2_{\bm{\mathcal{C}}_k} \\
    &= \frac{1}{2}[\bm{y}_k-H(\bm{x})]^\mathrm{T} \bm{R}^{-1}[\bm{y}_k-H(\bm{x})] + \frac{1}{2}[\bm{x}-\hat{\bm{x}}_k]^\mathrm{T}\bm{\mathcal{C}}_k^{-1}[\bm{x}-\hat{\bm{x}_k}],
\end{align}
where $\bm{R}$ is the measurement noise covariance matrix, which is an input to the algorithm, and $\bm{\mathcal{C}}_k$ is the ensemble covariance at time step $k$.
This covariance is defined as
\begin{gather}
    \bm{\mathcal{C}}_k = \bm{A}_k(\bm{A}_k)^\mathrm{T},
\end{gather}
where $\bm{A}_k$ is the state perturbation matrix
\begin{gather}
    \bm{A}_k = \frac{1}{\sqrt{q-1}}\left[\bm{x}_k^{(1)}-\overline{\bm{x}}_k, \ ... \ , \ \bm{x}_k^{(q)}- \overline{\bm{x}}_k\right].
\end{gather}
In fact, the minimization does not make use of the covariance matrix directly, but instead uses the state perturbation matrix and scaled output perturbation matrix $\bm{H}\bm{A}_k$ defined as
\begin{gather}
    \bm{H}\bm{A}_k = \frac{1}{\sqrt{q-1}}\left[\bm{y}_k^{(1)}-\overline{\bm{y}}_k, \ ... \ , \ \bm{y}_k^{(q)}- \overline{\bm{y}}_k\right].
\end{gather}

The optimization is carried out by finding the minimizer $\bm{x}_k$ satisfying
\begin{gather} \label{eq:update_ENKF}
    \bm{x}_k = \hat{\bm{x}}_k + \bm{A}_k \cdot \bm{w}_k,
\end{gather}
with $\bm{w}_k$ a correction coefficient.
This restricts the solution to the subspace spanned by the scaled perturbation matrix around the prior estimate $\hat{\bm{x}}_k$.
The optimization can be restated as
\begin{gather}
    \bm{w}_k = \underset{\bm{w} \in \mathbb{R}^q}{\mathrm{argmin}}~J(\bm{w}),
\end{gather}
where
\begin{gather}
    J(\bm{w})=\frac{1}{2}\lVert \bm{w}\rVert^2+\frac{1}{2}\lVert \bm{y}_k - H(\hat{\bm{x}}_k) - \bm{HA}_k(\bm{w})\rVert_{\bm{R}}^2.
\end{gather}
The solution is unique, and using the Woodbury matrix identity to write the inversion in measurement space, can be written as
\begin{gather} \label{eq:meas_space}
    \bm{w}_k = (\bm{HA}_k)^\mathrm{T}[\bm{R}+(\bm{HA}_k)(\bm{HA}_k)^\mathrm{T}]^{-1}(\bm{y}_k-H(\hat{\bm{x}}_k)).
\end{gather}
Performing this inversion in the measurement space is in most cases more computationally efficient.
Here this is clear, as the measurement space is comprised of only one variable (bubble radius).
Once the minimizer is found and the analysis step complete, covariance inflation is applied to the ensemble to correct for the (typical) underestimation of the variance with finite (typically small) ensembles (see section \ref{sec:cov_inf} for details on covariance inflation).
Finally, the forecast step can be repeated.

%%%%%

\subsection{The iterative ensemble Kalman smoother} \label{sec:IEnKS}

Minimizing deviation from data at future times can help to smooth out estimation and focus on longer-term trends.
The IEnKS uses information from one or multiple future time steps in its assimilation, and can thus be an effective tool.
While the ensemble initialization and forecast step are the same as that of the EnKF, the difference in the analysis step is twofold.
First, the cost function is modified to minimize difference with data at a single or multiple future times~\citep{Evensen_2000}.
The assimilation thus trails the simulation by a number of time steps (called the lag).
Second, it is no longer minimized analytically but iteratively using a Gauss--Newton algorithm.

The IEnKS method used here is from~\citet{Bocquet_2013} and~\citet{Sakov_2012}.
\citet{Bocquet_2013_2} have shown it to be effective for state and parameter estimation problems with highly nonlinear dynamics.
Its cost function can take two forms referred to as  `single data assimilation' (SDA) or `multiple data assimilation' (MDA)~\citep{Bocquet_2013}.
The IEnKS--SDA cost function penalizes difference with measurements at a single time step $k+L$, where $L$ corresponds to the lag of the smoother.
It is given by
\begin{gather}
    J(\bm{x}) = \frac{1}{2}\lVert \bm{y}_{k+L}-H\circ F_{k \rightarrow (k+L)}(\bm{x})\rVert^2_{\bm{R}} + \frac{1}{2}\lVert \bm{x}-\hat{\bm{x}}_k\rVert^2_{\bm{\mathcal{C}}_k}.
\end{gather}

On the other hand, the IEnKS--MDA cost function minimizes this difference over a data assimilation window (DAW) from $k+1$ to $k+L$, and is expressed as
\begin{gather}
    J(\bm{x}) = \frac{1}{2} \sum_{i=1}^L \beta_i \lVert \bm{y}_{k+i}-H\circ F_{k \rightarrow (k+i)}(\bm{x})\rVert^2_{\bm{R}} + \frac{1}{2}\lVert \bm{x}-\hat{\bm{x}}_k\rVert^2_{\bm{\mathcal{C}}_k},
\end{gather}
where $\beta_i$ are weights attributed to given time steps with $\sum\beta_i = 1$.
Again, a solution of the form $\bm{x} = \hat{\bm{x}}_k+\bm{A}_k \cdot \bm{w}$ is sought, but a Gauss--Newton method is used~\citep{Bocquet_2013}.
The minimizer $\bm{w}$ is found by iterating following
\begin{gather}
    \bm{w}_{(i+1)} = \bm{w}_{(i)} - \mathcal{H}_{(i)}^{-1}\Delta J_{(i)}(\bm{w}_{(i)}),
\end{gather}
where $i$ is the iteration number, and $\mathcal{H}$ is the approximate Hessian
\begin{gather}
    \mathcal{H}_{(j)} = (q-1)\bm{I} + \bm{H}\bm{A}_{(j)}^\mathrm{T} \bm{R}^{-1} \bm{H}\bm{A}_{(j)},
\end{gather}
where $\bm{I}$ is the $q \times q$ identity matrix.
The gradient is given by
\begin{gather}
    \Delta J_{(j)} = -\bm{H}\bm{A}_{(j)}^\mathrm{T}\bm{R}^{-1}[\bm{y}_{k+L} - H \circ F_{k+L \leftarrow k}(\bm{x}_k)]+(q-1)\bm{w}_{(j)}.
\end{gather}

For the smoother, $\bm{HA}$ is more complicated than it is for the filter as it involves differences with measurements at future time steps.
This quantity is akin to a tangent linear operator from ensemble to measurement space and has to be estimated.
Following~\citet{Bocquet_2013}, a finite-difference estimate is used:
\begin{gather}
    \bm{H}\bm{A}_{(j)} \approx \frac{1}{\alpha} H \circ F_{k+L \leftarrow k} (\bm{x}_k^{(j)} \bm{1}^\mathrm{T} + \alpha \bm{A}_k)\left(\bm{I}-\frac{\bm{1} \cdot \bm{1}^\mathrm{T}}{q}\right),
\end{gather}
with scaling factor $\alpha \ll 1$ and $\bm{1}=(1 \ \cdots \ 1)^\mathrm{T}$ a vector of length $q$.
The iteration is repeated until a threshold $\bm{w}_{(i+1)} - \bm{w}_{(i)} < \epsilon$, or a fixed number of iterations is reached.
Once the optimal value $\bm{w}_{\mathrm{opt}}$ is obtained, $\bm{x}_{\mathrm{opt}} = \hat{\bm{x}}+\bm{A}_k \cdot \bm{w}_{\mathrm{opt}}$ is calculated and a new ensemble $\bm{E}_k$ is sampled at time step $k$ with
\begin{gather}
    \bm{E}_k = \bm{x}_{\mathrm{opt}}\bm{1}^\mathrm{T} + \sqrt{q-1}\bm{A}_k\mathcal{H}_{\mathrm{opt}}^{-1/2}\bm{I}.
\end{gather}
This completes the analysis step.
When using the MDA variant, the Hessian and gradient of $J$ are found with
\begin{align}
    \mathcal{H}_{(j)} &= (q-1)\bm{I} + \sum^L_{i=1}\bm{HA}_i^\mathrm{T} \beta_i \bm{R}^{-1} \bm{HA}_i \label{eq:hess_mda}\\
    \Delta J_{(j)} &= -\sum^L_{i=1}\bm{HA}_i^\mathrm{T}\beta_i\bm{R}^{-1}[\bm{y}_{k+i} - H \circ F_{k+i \leftarrow k}(\bm{x}_i)]+(q-1)\bm{w}_{(j)}. \label{eq:deltaJ_mda}
\end{align}

%%%%%

\subsection{Covariance inflation} \label{sec:cov_inf}

While the EnKF and IEnKS may converge, ensemble methods are subject to intrinsic sampling error~\citep{Bocquet_2011, Luo_2011}.
This sampling error results from the finite ensemble size $q$ used to represent the statistics of a system of often much higher dimension.
As \citet{Leeuwen_1999} explains, the EnKF tends to underestimate error variances, particularly for small ensemble sizes.
There exist different ways to address this sampling error, but a simple approach is covariance inflation~\citep{Whitaker_2012}, where we correct 
\begin{gather} \label{eq:cov_inf}
    \bm{x}^{(j)} = \overline{\bm{x}} + \alpha(\bm{x}^{(j)} - \overline{\bm{x}}) + \bm{\lambda}^{(j)}.
\end{gather}
Here, $\overline{\bm{x}}$ denotes the ensemble average, as defined in~\eqref{eq:x_bar}, after the analysis step.
Parameters $\alpha$ and $\bm{\lambda}$ correspond to multiplicative and additive inflation parameters, respectively.

There exist many schemes for multiplicative inflation, the most simple of which is picking a scalar $\alpha$ (usually $1.005 \leq \alpha \leq 1.05$).
This can work well but requires extensive tuning to optimize the value for each run or data set.
Instead, \citet{Whitaker_2012} propose a scheme they call `Relaxation Prior to Spread' (RTPS).
Here, the value for $\alpha$ is found at each time step using
\begin{gather} \label{eq:RTPS}
    \alpha_i = 1 + \theta \left(\frac{\sigma_i^b - \sigma_i^a}{\sigma_i^a}\right),
\end{gather}
where $\sigma_i^a$ and $\sigma_i^b$ are the prior and posterior ensemble standard deviation for the $i^{th}$ element of the state vector ($\alpha$ is a vector here), and $\theta$ is a scalar (usually $0.5 \leq \theta \leq 0.95$).
As this expression for $\alpha$ shows, this scheme inflates the covariance more in regions where the analysis led to a large correction.
\citet{Whitaker_2012} test this method and compare it to other approaches, showing that it performs well.
We similarly find that this performs as well or better than a simple scalar $\alpha$ for our tested cases.
This RTPS model was used with $\theta = 0.7$.
Additive covariance inflation was not found to significantly affect results and introduced some stability issues with larger magnitudes of $\bm{\lambda}$.
Therefore, $\bm{\lambda} = \bm{0}$ is used.

%%%%%

\subsection{A hybrid ensemble-based 4D--var method} \label{sec:En4D-Var}

As with ensemble Kalman methods, ensembles can be used with 4D-Var to estimate covariance empirically, thus reducing computational cost~\citep{Gustafsson_2014,Liu_2008}.
The present method (En4D-Var) is a fully offline extension of the IEnKS--MDA method.
Again, the ensemble is initialized in the same way as EnKF, but the cost function is here
\begin{gather} \label{eq:cost_En4D}
    J(\bm{x}) = \frac{1}{2} \sum_k \beta_k 
    \lVert \bm{y}_{k}-H\circ F_{k \leftarrow 0}(\bm{x})\rVert^2_{\bm{R}} + 
    \frac{1}{2} \lVert \bm{x}-\hat{\bm{x}}_k\rVert^2_{\bm{\mathcal{C}}_0}.
\end{gather}
The difference with the IEnKS--MDA cost function is the data assimilation window size.
Rather than minimizing over a few time steps forward and then stepping through time, the minimization is done over the entire time domain and only the initial state vector is corrected.
Each new iteration is initialized with the corrected initial state (including parameters to estimate).
The same minimization procedure as described in section~\ref{sec:IEnKS} is used.
When the minimization has converged, a final simulation is run with the forecast model only.
In cases where only a few iterations are necessary, this method reduces computational cost as compared to the IEnKS-MDA.
The time dimension is still full included, but each point in time is only assimilated once per iteration.
Furthermore, this retains the advantage of ensemble methods.
As opposed to classical 4D--Var, there is no need to linearize the state function and find the tangent linear adjoint operator.
This novel adaptation of the IEnKS method is well suited to the present problems given that our interest is the estimation of material properties which are, at the outset, assumed to be constant.

%%%%%%%%%%%%%%%%%%%%%%%%%%%%%%
%%% Synthetic Data Results %%%
%%%%%%%%%%%%%%%%%%%%%%%%%%%%%%

\section{Testing with synthetic data} \label{sec:results_sim}

%%%%%

\subsection{Parameter estimation results} \label{sec:sim_perf}

Synthetic data where the true shear modulus and viscosity are known is generated from the model (section~\ref{sec:physical_model}) and used to test the data assimilation methods in a setting where there is no modeling error.
Bubble radius time-history data from the simulation is sampled at 270,000 frames per second to match available experiments.
Random Gaussian noise is added to these samples to mimic experimental data.
The standard deviation of this noise is set at $\sigma = 0.02$, which is greater than the estimated noise of the experiments.
Two polyacrylamide gels were examined with nominal values of shear modulus and viscosity determined by \citet{Estrada_2018}.
For the stiff gel: $G_{\mathrm{stiff}} = \SI{7.69}{kPa}$,\ $\mu_{\mathrm{stiff}} = \SI{0.101}{Pa.s}$, and for the soft gel: $G_{\mathrm{soft}} = \SI{2.12}{kPa}$,\ $\mu_{\mathrm{soft}} = \SI{0.118}{Pa.s}$.
Since similar estimation accuracy was achieved in both cases, we report results for the stiff gel only.
The other material properties used are taken from \citet{Estrada_2018} and given in table \ref{tab:material_params}.
No uncertainty is added to these parameters in the present study to match their conditions and focus on estimating $G$ and $\mu$.

\begin{table}[H]
    \centering
    \begin{tabular}{|c|c||c|c|}
        \hline 
        \textbf{Parameter} & \textbf{Value} & \textbf{Parameter} & \textbf{Value} \\
        \hline
        $\rho$ & $\SI{1060}{kg/m^3}$ & $c$ & $\SI{1430}{m/s}$ \\
        $p_{\infty}$ & $\SI{101.3}{kPa}$ & $\gamma$ & $\SI{5.6e-2}{N/m}$\\
        $D$ & $\SI{24.2e-6}{m^2/s}$ & $\kappa$ & $1.4$ \\
        $C_{p,g}$ & $\SI{1.62}{kJ/kg.K}$ & $C_{p,v}$ & $\SI{1.00}{kJ/kg.K}$ \\
        $A$ & $\SI{5.3e-5}{W/m.K^2}$ & $B$ & $\SI{1.17e-2}{W/m.K^2}$ \\
        $p_{\mathrm{ref}}$ & $\SI{1.17e8}{kPa}$ & $T_{\mathrm{ref}}$ & $\SI{5200}{K}$ \\
        $T_{\infty}$ & $\SI{298.15}{K}$ & & \\
        \hline
    \end{tabular}
    \caption{Model parameters as they follow from~\citet{Estrada_2018}.}
    \label{tab:material_params}
\end{table}

\begin{figure}[H]
\centering
\begin{subfigure}{.5\linewidth}
    \centering
    \includegraphics[width=\linewidth]{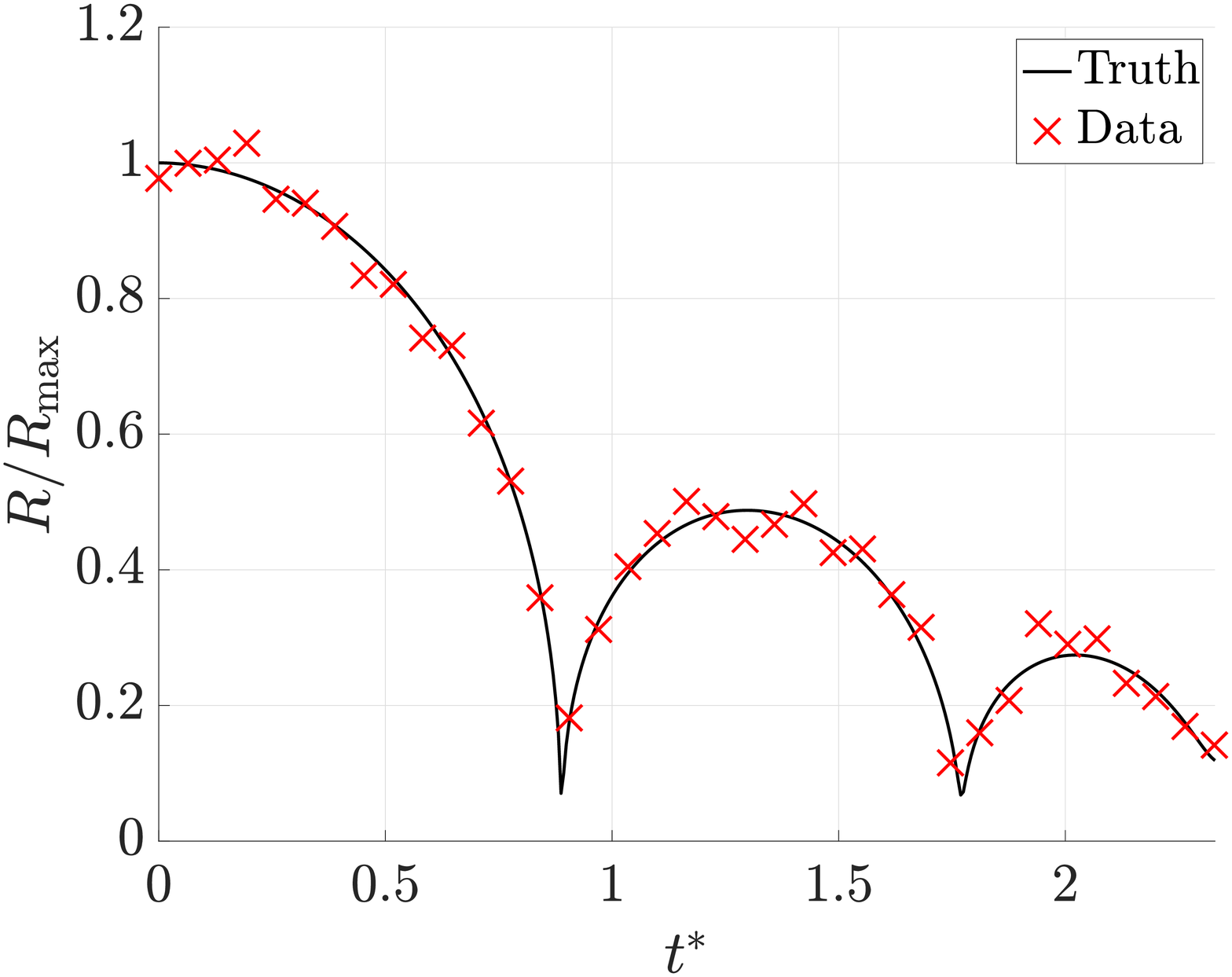}
    \subcaption{}
    \label{fig:sim_radius}
    \end{subfigure}%
    \begin{subfigure}{.5\linewidth}
    \centering
    \includegraphics[width=\linewidth]{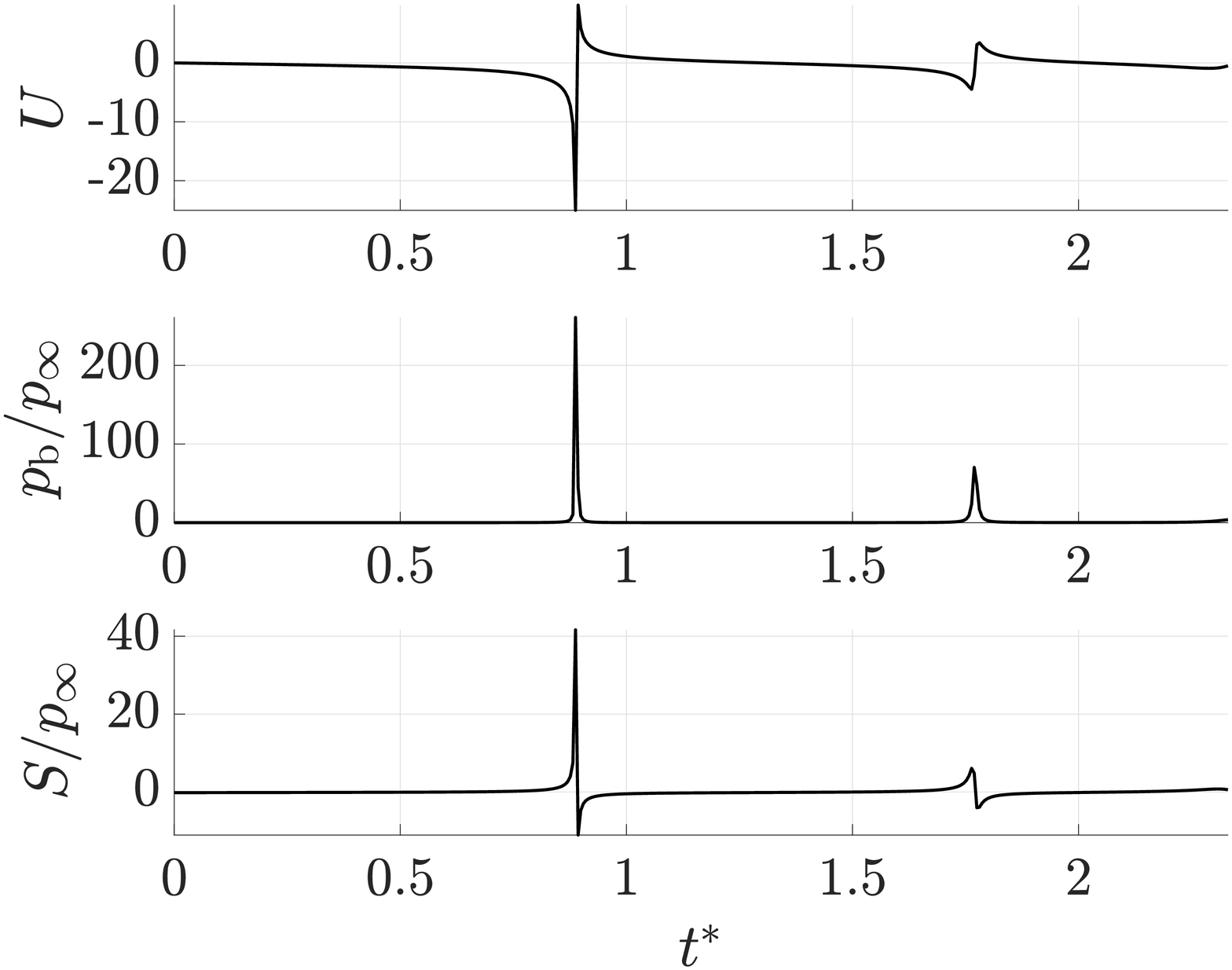}
    \subcaption{}
    \label{fig:sim_UPS}
    \end{subfigure}
    \caption{Simulated bubble radius and noisy sampled data used to test data assimilation methods \textbf{(\subref{fig:sim_radius})}, alongside simulated bubble-wall velocity, normalized bubble pressure and stress integral \textbf{(\subref{fig:sim_UPS})}, plotted over non-dimensional time $t^\ast$.}
    \label{fig:sim_run}
\end{figure}

An example simulated radius curve and sampled surrogate measurements (with noise added) with these parameters is shown in figure~\ref{fig:sim_radius}, plotted against non-dimensional time
\begin{equation}
    t^\ast = \frac{t}{R_{\mathrm{max}}}\sqrt{\frac{p_\infty}{\rho}}.
\end{equation}
With the simulated data, the evolution over time of all variables in the state vector is known.
For example, bubble-wall velocity, bubble pressure and stress integral are plotted in figure~\ref{fig:sim_UPS}.

A set of initial guesses for the shear modulus and viscosity, ranging from 10\% to 100\% error from the true values, were used to test each method.
Table \ref{tab:combined} summarizes results for a subset of these cases, representing 10, 50 and 100\% initial error in $G$ and $\mu$.
In each case, ensembles were initialized as Gaussian with these erroneous material properties as the mean, and standard deviation increasing with increased error.
That is, the spread of the initial ensemble was made wider for cases with more error, to account for the increased uncertainty in the initial guess.
To match the tests on experimental data in the next section, simulated data is limited to the first three peaks of the bubble collapse.
This corresponds to approximately 35 points given the initial conditions and frame rate.
\citet{Estrada_2018} found that limiting the data to this region led to better parameter estimation.
Similarly, we find that the model fails to fit the radius measurements after this time.
Reasons for this reduced accuracy at later times are discussed in section~\ref{sec:exp_discussion}. \\

\begin{table}[H]
    \centering
    \begin{tabular}{|c|c|c|c|}
    \hline
    Method & $G$ estimate (\%error) [\si{kPa}] & $\mu$ estimate (\%error) [\si{Pa.s}]& Run time [\si{s}]\\
    \hline\hline
     Guess 1 & $8.50$ (+10\%)  & $0.09$ (-10\%)  & -- \\
    \hline
        EnKF & $7.234$ (5.93\%) & $0.098$ (2.61\%) & 428\\
        IEnKS--SDA (lag 1) & $7.364$ (4.24\%) & $0.110$ (8.58\%) & 852\\
        IEnKS--MDA (lag 3) & $6.682$ (13.11\%) & $0.100$ (0.92\%) & 8076\\
        En4D--Var & $7.150$ (7.03\%) & $0.099$ (1.80\%) & 679\\
        \hline
        
       \hline\hline
     Guess 2 & $3.80$ (-50\%)  & $0.05$ (-50\%)   & -- \\
    \hline
         EnKF & $3.988$ (48.1\%) & $0.057$ (43.1\%) & 375\\
        IEnKS--SDA (lag 1) & $4.203$ (45.4\%) & $0.080$ (20.9\%) & 904\\
        IEnKS--MDA (lag 3) & $7.390$ (3.90\%) & $0.086$ (15.2\%) & 9755\\
        En4D--Var & $7.396$ (3.82\%) & $0.100$ (0.52\%) & 690\\
        \hline
  
     \hline\hline
     Guess 3 & $15.0$ (+100\%)  & $0.20$ (+100\%)  & -- \\
    \hline
        EnKF & $13.649$ (77.5\%) & $0.175$ (73.1\%) & 495\\
        IEnKS--SDA (lag 1) & $10.272$ (33.6\%) & $0.142$ (40.7\%) & 800\\
        IEnKS--MDA (lag 3) & $10.078$ (31.1\%) & $0.121$ (19.9\%) & 9802\\
        En4D--Var & $10.210$ (32.7\%) & $0.118$ (16.6\%) & 611\\
        \hline

    \end{tabular}
    \caption{Comparing accuracy of estimation with 3 different initial guesses for the parameters. Runs were performed on a machine with dual 12-core 2.3Ghz processors}
    \label{tab:combined}
\end{table}

Table \ref{tab:combined} shows that with a relatively good initial guess with $10\%$ error, the assimilation methods perform adequately.
For example, the EnKF tracks the correct values for shear modulus and viscosity within $6\%$ and $3\%$ respectively.
With a moderate initial error of $50$\%, however, the EnKF looses accuracy and barely improves on the initial guess.
In some cases, the EnKF was observed to be unstable, and the initial ensemble covariance had to be limited to prevent divergence.
This limited the ability of the filter to estimate the parameters of interest, and thus despite its computational efficiency, the EnKF is eliminated from further consideration.

The IEnKS and En4D--Var performed better than the EnKF for the $50$\% error case.
The estimation was stable while varying initial conditions and covariance.
Still, the lag 1 IEnKS--SDA only resulted in marginal improvements in the parameter values.
The lag 3 IEnKS--MDA, on the other hand, resulted in further improvement, but at a high computational cost.
This cost is associated with the calculation of the Hessian (see equation~\eqref{eq:hess_mda}) and gradient of the cost function (see equation~\eqref{eq:deltaJ_mda}), which now involves three future time steps.
The En4D--Var performs best in this test case, achieving good estimation with a comparably fast computational time.
We note that while the En4D--Var was run for fifteen iterations in each case, the material property estimation converged by the fifth iterations.
Thus, results and run time after five iterations are reported.

\begin{figure}[H]
\centering
    \begin{subfigure}{.5\linewidth}
        \centering
        \includegraphics[width=\linewidth]{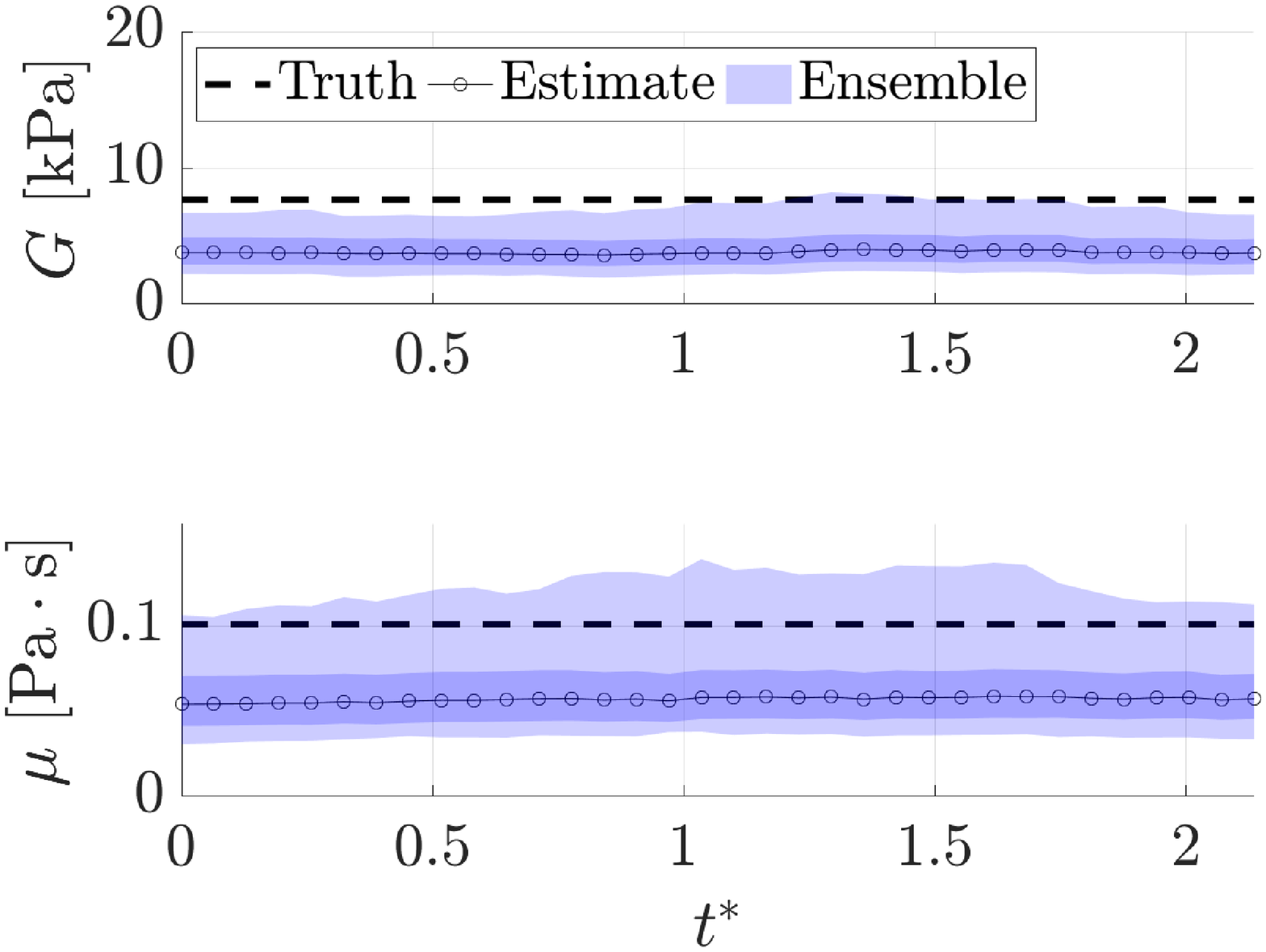}
        \caption{EnKF}
        \label{fig:sim_EnKF_50_params}
    \end{subfigure}%
    \begin{subfigure}{.5\linewidth}
        \centering
        \includegraphics[width=\linewidth]{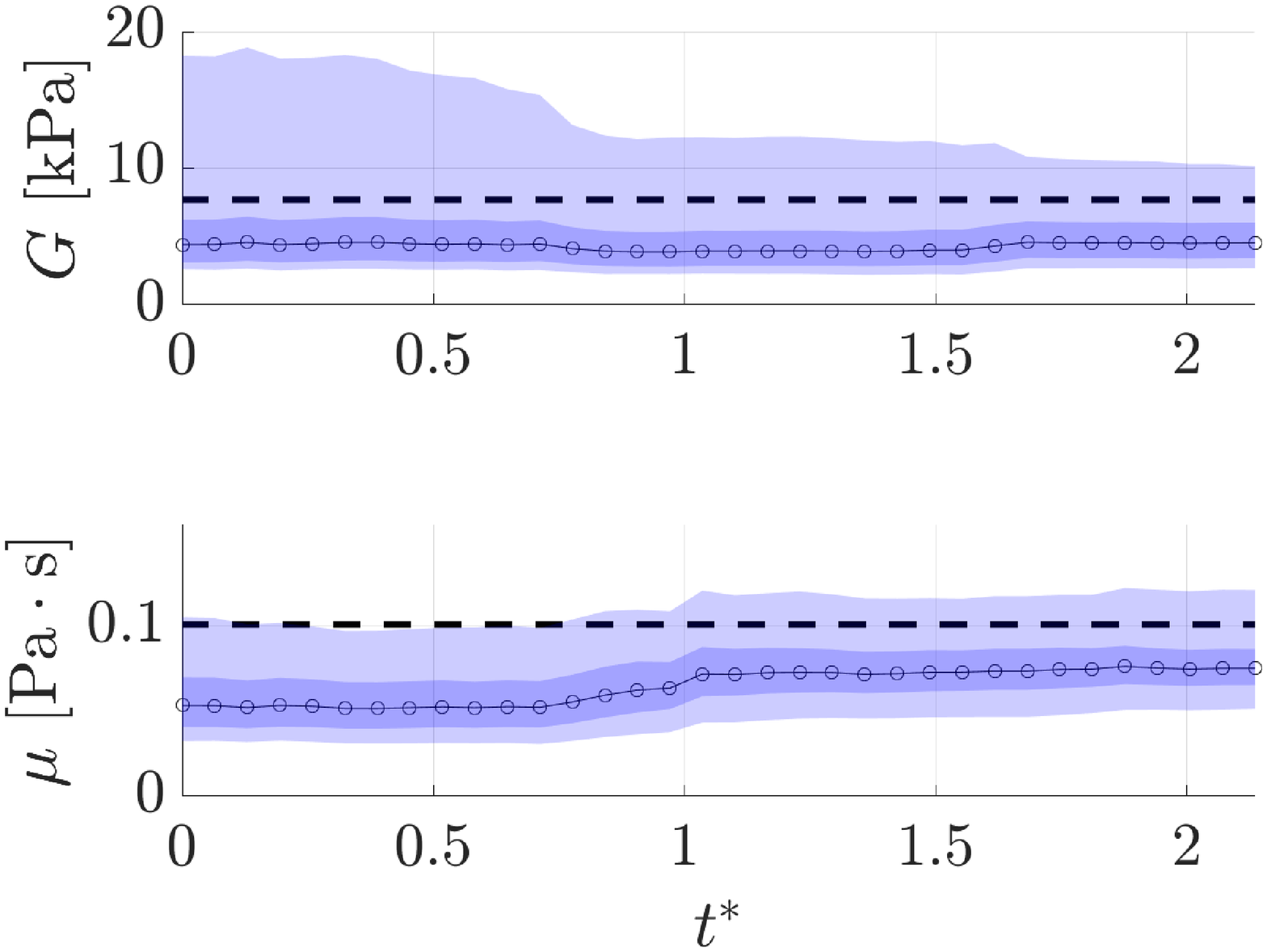}
        \caption{IEnKS--SDA (lag 1)}
        \label{fig:sim_sda_50_params}
    \end{subfigure} \\
    \begin{subfigure}{.5\linewidth}
        \centering
        \includegraphics[width=\linewidth]{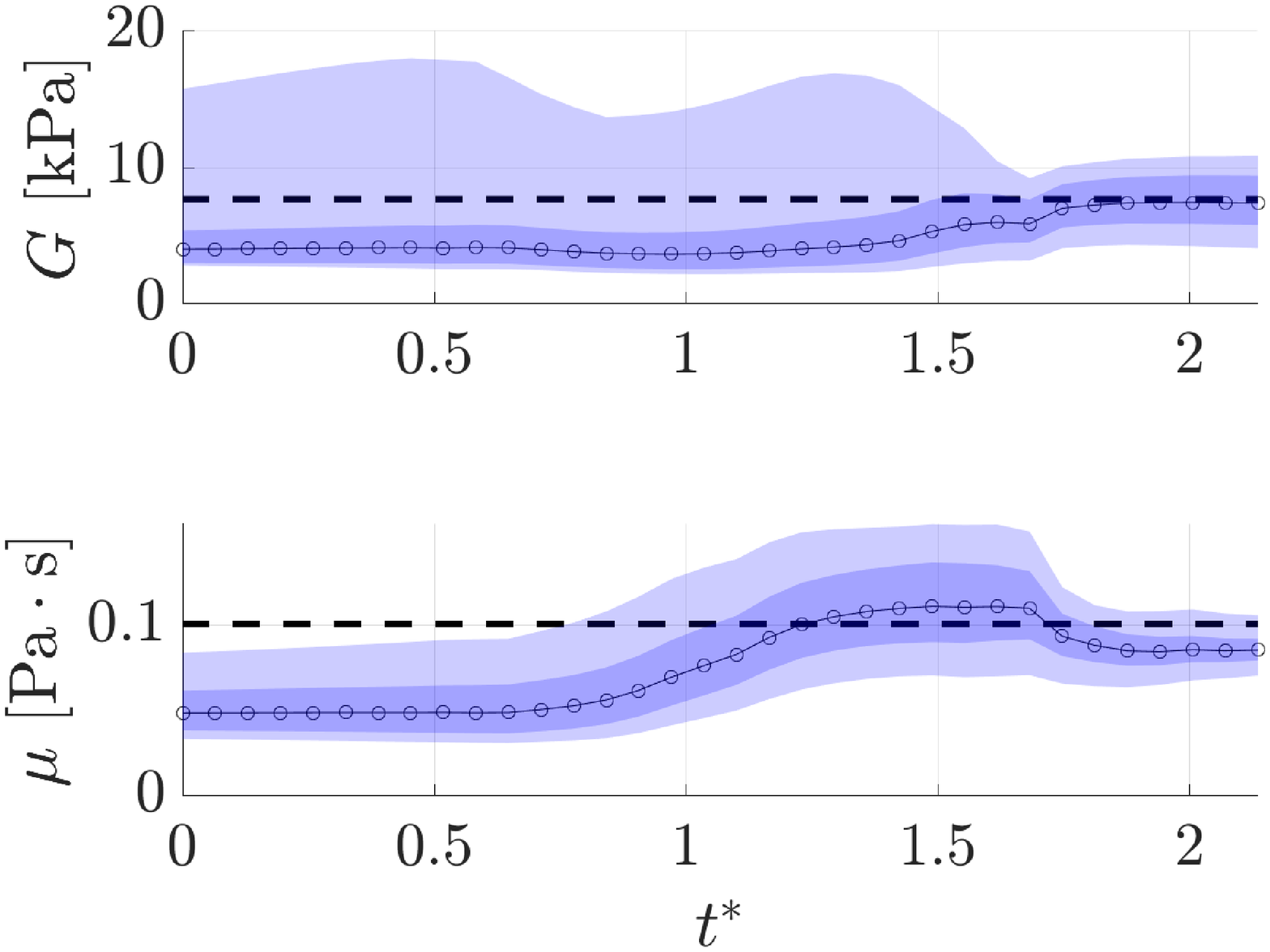}
        \caption{IEnKS--MDA (lag 3)}
        \label{fig:sim_mda_50_params}
    \end{subfigure}%
    \begin{subfigure}{.5\linewidth}
        \centering
        \includegraphics[width=\linewidth]{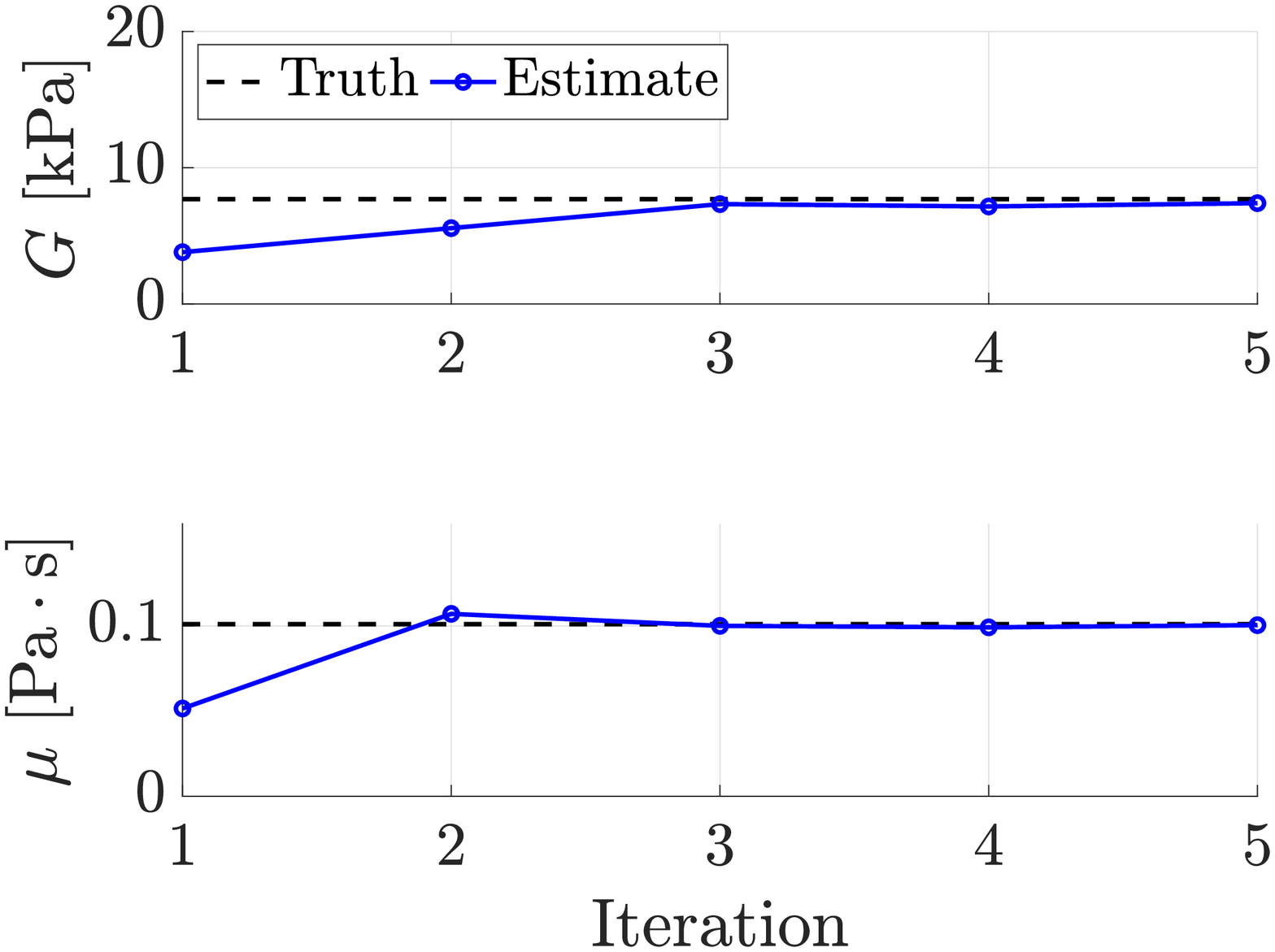}
        \caption{En4D--Var}
        \label{fig:sim_En4D_50_params}
    \end{subfigure}
\caption{Estimation of shear modulus and viscosity with initial guesses of $G=\SI{3.8}{kPa}$ and $\mu =\SI{0.05}{Pa.s}$ (both at 50\% error). The estimation is plotted over non-dimensional time $t$ for the EnKF and IEnKS methods, and over iteration number for the En4D--Var.}
\label{fig:sim_50}
\end{figure}

Estimation results in the case with 50\% error are presented in figure~\ref{fig:sim_50}.
Figure~\ref{fig:sim_En4D_50_params} shows the suitability of the En4D--Var: both parameters converge to accurate estimates within a few iterations.
Overall, figure~\ref{fig:sim_50} also highlights the value of looking over a time horizon.
While the EnKF and lag 1 IEnKS appear to disbelieve the data too much throughout the run, taking into account multiple times enables the lag 3 IEnKS--MDA to adjust to new information well, notably around collapse.
Indeed, the IEnKS--MDA significantly corrects the viscosity estimate around each collapse, and the shear modulus estimate during the second collapse.
Assimilating data from single time-steps appears to be insufficient given the short time scales of bubble cavitation and limited data.
Smoothing over multiple times far improves performance around collapse points, which, given the IEnKS--MDA results, appear to hold the most pertinent data to make the necessary corrections.

In the case with $100$\% error in the initial guess, the relative performances of each method are similar to the $50$\% error case, but the three smoothers stagnate at $20$ to $40$\% errors for $\mu$ and $G$.
Weighted by computational expense, the En4D--Var performs best, but the IEnKS--MDA should not be discarded.
Indeed, the time-varying estimation provides additional information about potentially time-dependent modeling uncertainties.
While the physical model used assumes a constant shear modulus and viscosity, the quasi-online IEnKS--MDA can uncover potential limitations of this assumption.
This issue is further examined in section~\ref{sec:exp_discussion}.

%%%%%

\subsection{Uncertainty} \label{sec:sim_uncertainty}

\begin{figure}[t]
    \centering
    \includegraphics[width=.5\textwidth]{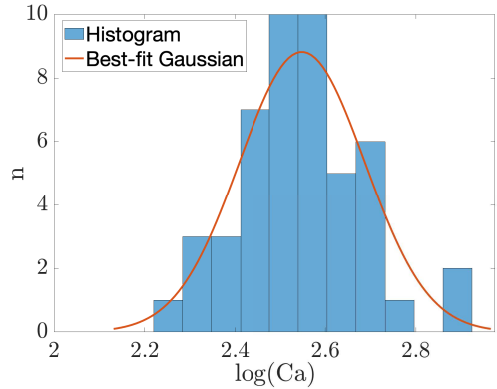}
    \caption{Histogram of the final estimate for $\log(\Ca)$ with the lag 1 IEnKS and fitted normal curve, where n is the number of ensemble members at each value of $\log(\Ca)$.}
    \label{fig:logCa_histogram}
\end{figure}

Ensemble methods carry information about error statistics of the estimated parameters in the final ensemble.
One way to visualize ensembles is through a histogram, an example of which is shown in figure~\ref{fig:logCa_histogram} for the logarithm of the Cauchy number with the lag 1 IEnKS--SDA estimator.
Despite the nonlinearity of the model, the tested methods track only the first two statistical moments of an assumed Gaussian filtering or smoothing PDF.
Previous works (e.g.,~\citep{Evensen_2000, Yang_2012, Katzfuss_2016}) have discussed that adequate results can still be achieved with a nonlinear model where this assumption must break down to some degree.
Our results for the IEnKS and En4D--Var results above confirm that this is the case in this example.

\begin{figure}[t]
    \centering
    \begin{subfigure}{.5\textwidth}
    \centering
    \includegraphics[width=\textwidth]{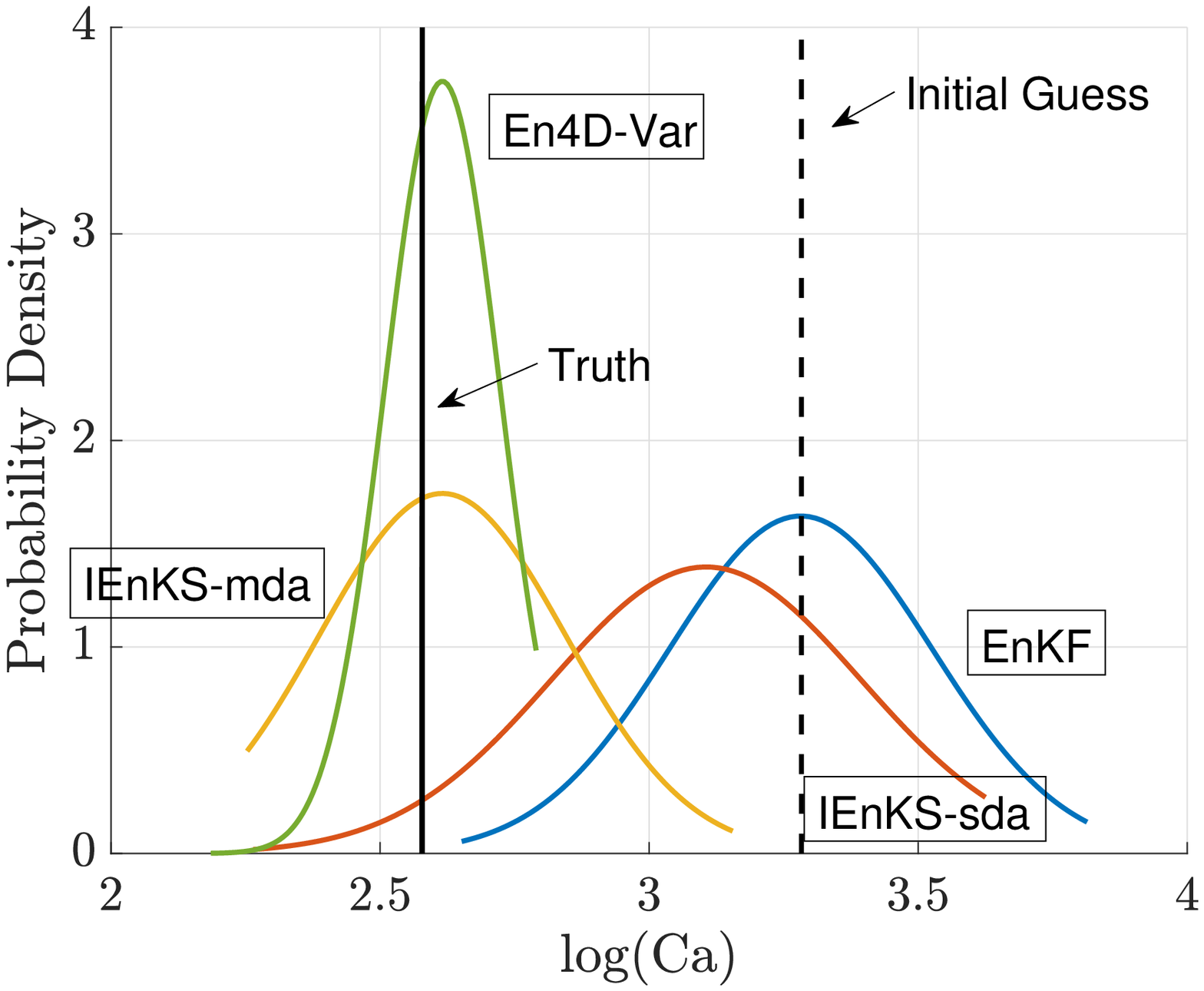}
    \caption{}
    \label{fig:log_Ca_distrib_case2}
    \end{subfigure}%
    \begin{subfigure}{.5\textwidth}
    \centering
    \includegraphics[width=\textwidth]{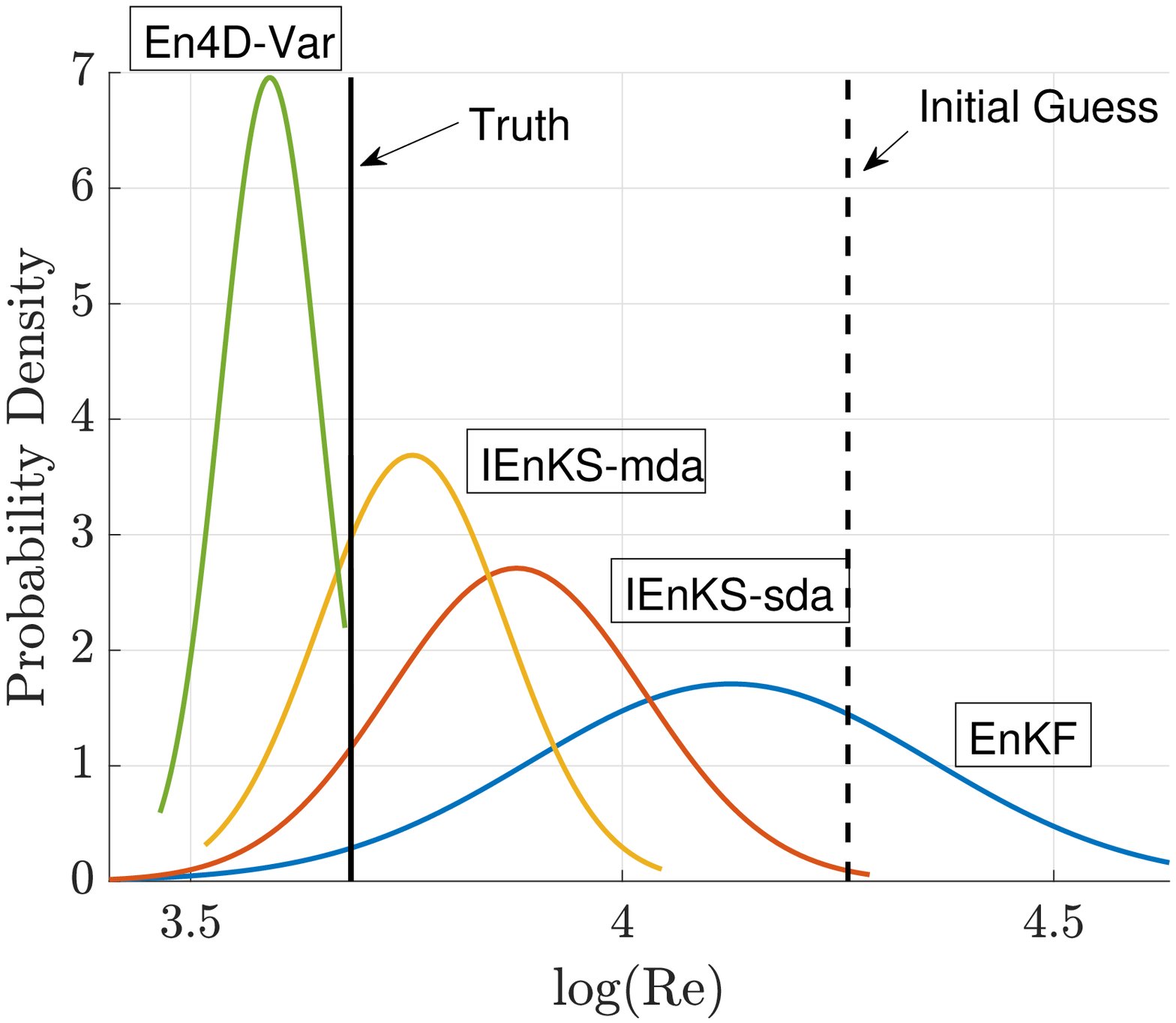}
    \caption{}
    \label{fig:log_Re_distrib_case2}
    \end{subfigure}
    \caption{Comparing final ensembles for log(Ca) \textbf{(\subref{fig:log_Ca_distrib_case2})} and log(Re) \textbf{(\subref{fig:log_Re_distrib_case2})} in the case with 50\% initial error in both parameters.}
    \label{fig:distrib_case2}
\end{figure}

Figure~\ref{fig:distrib_case2} shows a comparison of the fitted histograms for the methods for the case with $50\%$ initial error in both parameters.
Despite imperfect estimation, the En4D--Var converges significantly more than other methods given the limited data.
The IEnKS--MDA curve displays the least variance of the Kalman methods, as expected.

\begin{figure}[H]
    \centering
    \begin{subfigure}{.5\textwidth}
    \centering
    \includegraphics[width =\textwidth]{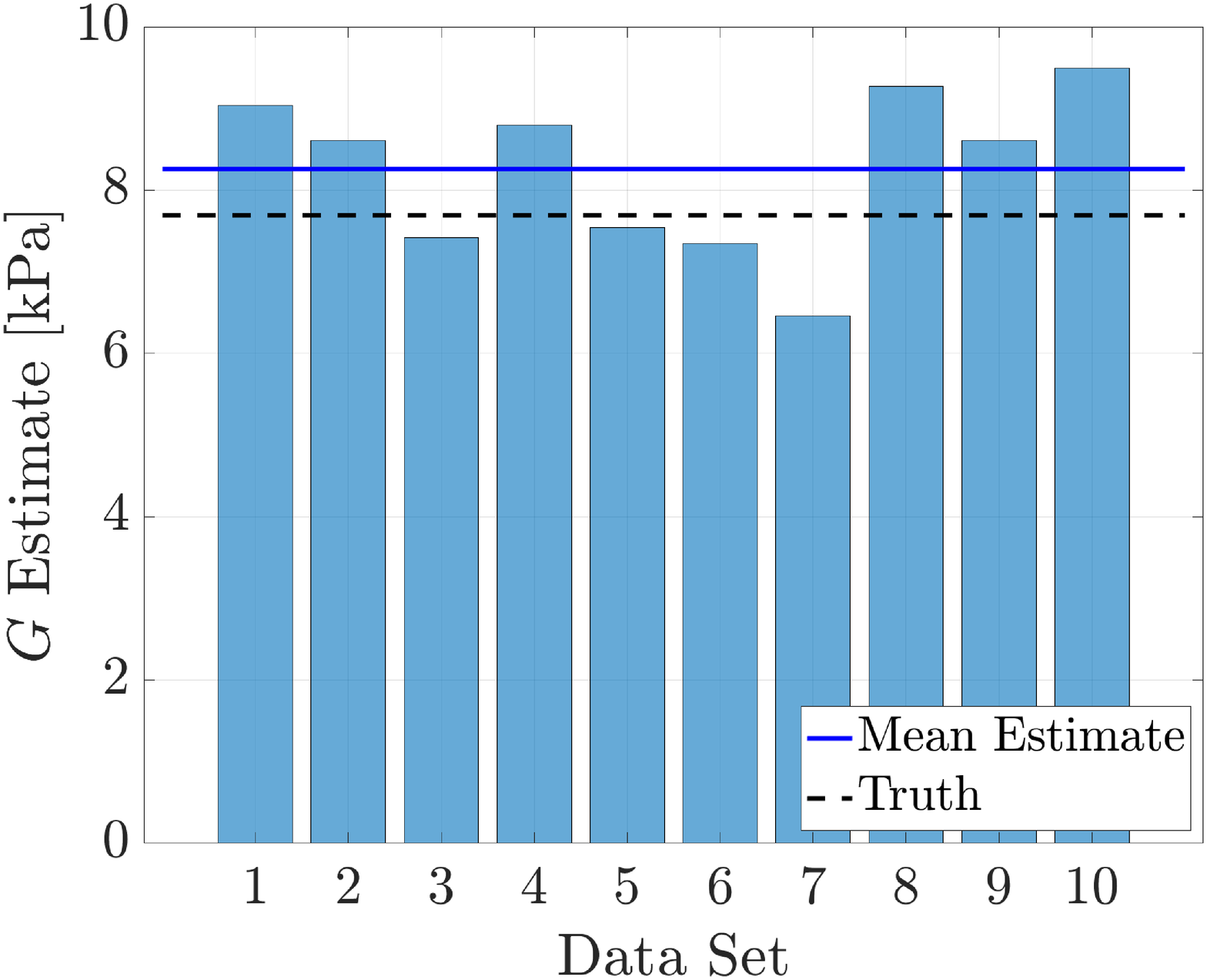}
    \caption{}
    \label{fig:sim_barplot_G}
    \end{subfigure}%
    \begin{subfigure}{.5\textwidth}
    \centering
    \includegraphics[width =\textwidth]{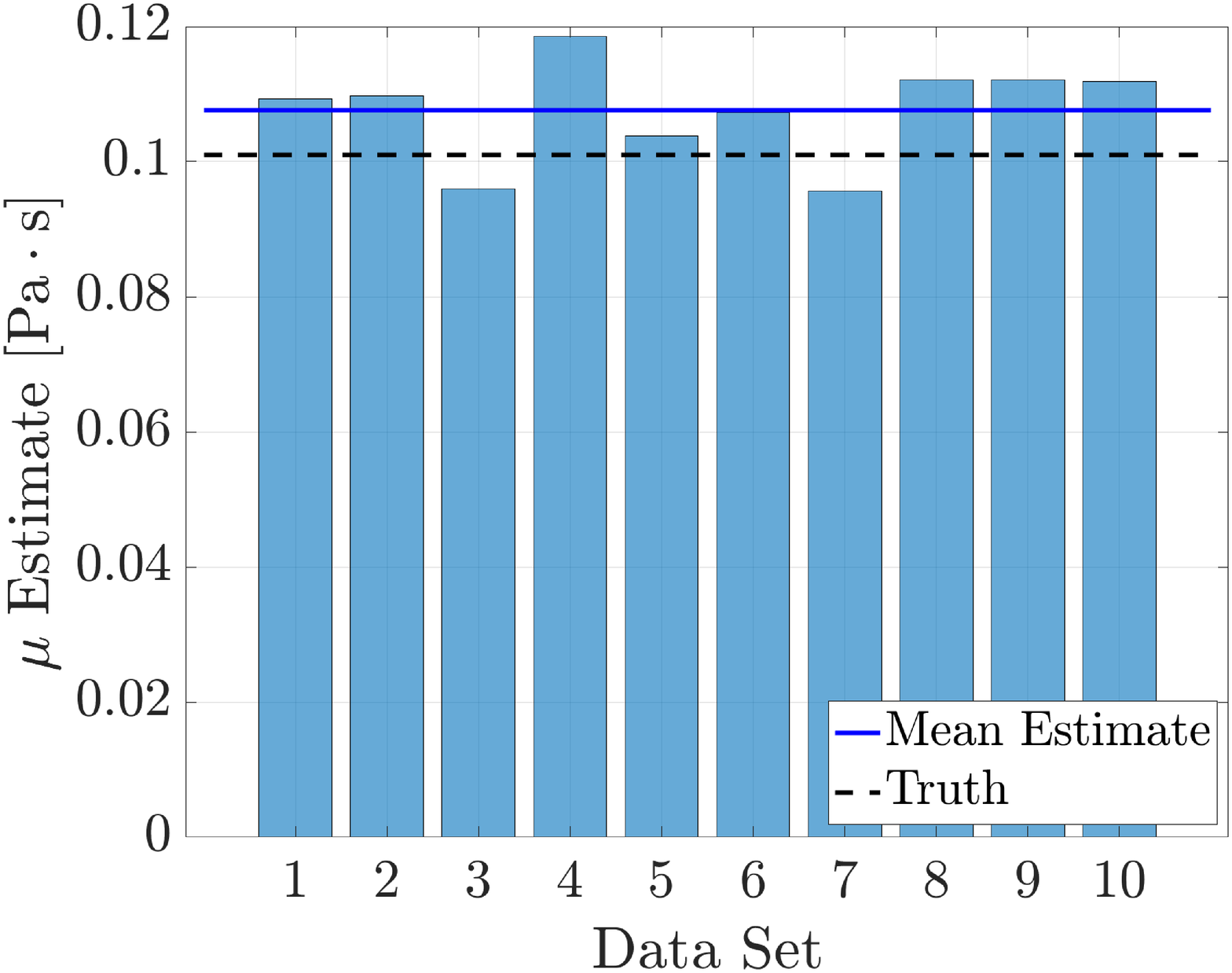}
    \caption{}
    \label{fig:sim_barplot_mu}
    \end{subfigure}
    \caption{En4D--Var results for G \textbf{(\subref{fig:sim_barplot_G})} and $\mu$ \textbf{(\subref{fig:sim_barplot_mu})}, for ten simulated data sets.}
    \label{fig:sim_bar_plots}
\end{figure}

\begin{figure}[H]
    \centering
    \includegraphics[width = .5\textwidth]{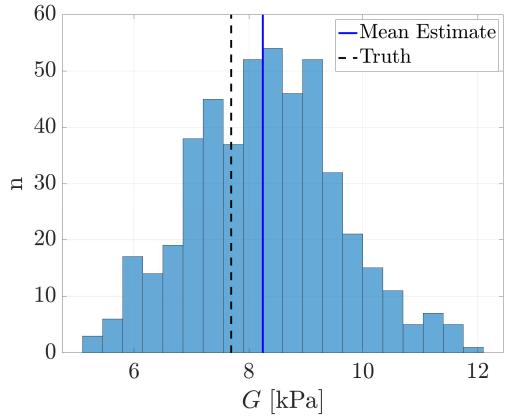}
    \caption{Histogram for G combining 10 final ensembles for simulated data runs with En4D-Var.}
    \label{fig:sim_hist_G}
\end{figure}

Anticipating experimental results, the En4D--Var was run for 10 simulated data sets with the same ground truth but different (random) noise.
Results are shown in figure~\ref{fig:sim_bar_plots} for shear modulus and viscosity estimates over the data sets.
The dashed black lines correspond to the truth, and the blue line to the mean estimate over the 10 runs.
Results across these 10 data sets are fairly uniform (standard deviation of $\SI{0.98}{kPa}$ for $G$, $\SI{0.009}{Pa.s}$ for $\mu$), confirming that reliable estimates are obtained despite noisy measurements across data sets.
Figure~\ref{fig:sim_hist_G} shows a histogram combining final ensembles for shear modulus to visualize overall results.
As each of the 10 ensembles should be approximately normal, a Gaussian curve is expected when combining them.
Figure~\ref{fig:sim_hist_G} indeed shows an approximately normal distribution, as does the equivalent histogram for viscosity (as shown in section~\ref{sec:exp_discussion} in figure~\ref{fig:sim_hist_mu}).

Based on these results with simulated data, given reasonable initial guesses as to the shear modulus and viscosity, we can confidently expect to estimate both parameters to within $5\%$ using the 10 available data sets.
Multiple initial guesses can be tested and their fits with experimental radius histories compared.
Therefore, it is straightforward to formulate an initial guess with less than $50\%$ error, and thus obtain results comparable to the second test case with simulated data.
The En4D--Var is the baseline given its performance.
IEnKS--MDA is also tested for its ability to estimate parameters quasi-online.

%%%%%%%%%%%%%%%%%%%%%%%%%%%%%%
% Experimental Data Results %%
%%%%%%%%%%%%%%%%%%%%%%%%%%%%%%

\section{Applied to experimental data} \label{sec:exp}

%%%%%

\subsection{Experimental setup}

For a more detailed description of the experimental setup for data collection, see \citet{Estrada_2018}.
After the polyacrylamide gel is prepared, each cavitation event is induced with a 6ns pulse of a ``user-adjustable 1--50 \si{mJ}, frequency-doubled Q-switched \SI{532}{\nano\meter} Nd:YAG laser".
These cavitation events are triggered at different locations in the same large batch of polyacrylamide to maximize uniformity of material properties across experiments.
Bubble radius is captured approximately every $\SI{3.7}{\micro\second}$, processing $\SI{270000}{fps}$ high-speed camera output by subtracting a reference image from each frame and fitting a circle.
A few sources of error may be present.
Nonuniformity of the polyacrylamide gel or discrepancies across data sets could cause the bubble to lose spherical symmetry.
Laser pulses may also vary slightly across runs, affecting the energy deposited in the system and thus initial growth conditions.
In practice, a difference in maximum bubble radii was observed, with $R_{\mathrm{max}} = \SI{388 \pm 35}{\micro\meter}$ across experiments with the stiff gel, and $R_{\mathrm{max}} = \SI{430 \pm 17}{\micro\meter}$ with the soft gel.
Ten experimental data sets in the stiff gel were used for the following results, in part to address this potential lack of uniformity in experimental conditions.

%%%%%

\subsection{Initial estimates for $G$ and $\mu$ with En4D--Var} \label{sec:exp_results}

The noise magnitude in the experimental data is smaller than what was used in the simulated data with the same data rate.
Therefore, if the model is adequate and the noise accurately represented as Gaussian, the IEnKS and En4D--Var should yield comparable or better estimation results with the experimental data.
As with the simulated data, the assimilation window is limited to the initial collapse and two subsequent rebounds to match the setup used by \citet{Estrada_2018}, who estimated in the stiff-gel: $G_{\mathrm{stiff}}=\SI{7.69 \pm 1.12}{kPa}$ and $\mu_{\mathrm{stiff}} = \SI{0.101 \pm 0.023}{Pa.s}$.
The results are compared to theirs.
Our estimation is initialized with three different initial guesses, detailed in table \ref{tab:exp_combined}.
Similarly to the surrogate truth data from the last section, initial guesses with  10\%, 50\% and 100\% difference from the \citet{Estrada_2018} estimates are chosen.

\begin{table}[H]
    \centering
    \begin{tabular}{|c|c|c|c|}
    \hline
    Method & $G$ estimate $\pm \sigma$ [\si{kPa}]& $\mu$ estimate $\pm \sigma$
    [\si{Pa.s}]& Run time [\si{s}]\\
    \hline\hline
     Guess 1 & $8.50$ (+10\% diff)  & $0.09$ (-10\% diff)  & -- \\
    \hline
        IEnKS--SDA (lag 1) & $7.93 \pm 1.68$ & $0.096 \pm 0.012$ & 2751 \\
        IEnKS--MDA (lag 3) & $7.51 \pm 1.50$ & $0.089 \pm 0.016$ & 9536 \\
        En4D--Var & $7.41 \pm 1.63$ & $0.093 \pm 0.014$ & 609 \\
        \hline
        
       \hline\hline
     Guess 2 & $3.80$ (-50\% diff)  & $0.05$ (-50\% diff)   & -- \\
    \hline
        IEnKS--SDA (lag 1) & $4.32 \pm 0.46$ & $0.085 \pm 0.013$ & 2832 \\
        IEnKS--MDA (lag 3) & $6.67 \pm 1.43$ & $0.083 \pm 0.016$ & 10052 \\
        En4D--Var & $6.53 \pm 1.58$ & $0.090 \pm 0.014$ & 585 \\
        \hline
  
     \hline\hline
     Guess 3 & $15.0$ (+100\% diff)  & $0.20$ (+100\% diff)  & -- \\
    \hline
        IEnKS--SDA (lag 1) & $9.46 \pm 2.76$ & $0.114 \pm 0.014$ & 2871 \\
        IEnKS--MDA (lag 3) & $8.57 \pm 1.52$ & $0.103 \pm 0.015$ & 9222 \\
        En4D--Var & $8.24 \pm 1.58$ & $0.098 \pm 0.016$ & 535 \\
        \hline

    \end{tabular}
    \caption{Comparing results of estimation with three different initial guesses for the parameters. Runs were again performed on a machine with dual 12-core 2.3Ghz processors}
    \label{tab:exp_combined}
\end{table}

Table \ref{tab:exp_combined} summarizes the results from the three different initial material parameters guesses for the three methods.
These estimates correspond to the mean estimate over all 10 experimental data sets.
The standard deviation $\sigma$ of the results is also reported.

\begin{figure}[H]
    \centering
    \includegraphics[width=.5\textwidth]{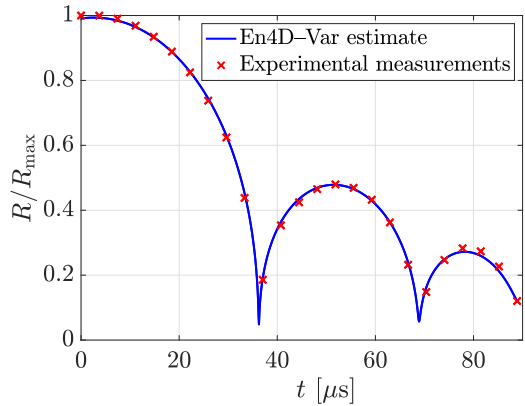}
    \caption{Radius curve given by En4D--Var estimates and experimental measurements for data set 10.}
    \label{fig:exp_En4D_fit}
\end{figure}

The En4D--Var estimates for shear modulus and viscosity all fall within the error bounds provided by \citet{Estrada_2018}.
Results are close to theirs in the test cases considered.
The estimates are uniform, with only $8.5\%$ difference between viscosity estimates, and $23\%$ difference in the shear modulus results across the data sets.
This larger difference in the shear modulus estimates and in the associated standard deviations is expected, as we have found the radius curves to be relatively more sensitive to $\mu$ than $G$.
Finally, the average normalized root mean squared error (NRMSE) for bubble radius is low at $\num{2.16e-2}$ for guess 1, indicating that a good fit was achieved with this method (see equation~\eqref{eq:nrmse} for NRMSE definition).
Given that the guess--1 results lead to the smallest radius error, our initial shear and viscosity modulus estimates are $G = \SI{7.41 \pm 1.63}{kPa}$ and $\mu = \SI{0.093 \pm 0.014}{Pa.s}$, respectively.
An example bubble radius curve is shown in figure~\ref{fig:exp_En4D_fit}, for one of the experimental data sets (data set 10).

The standard deviation across the 10 IEnKS--SDA runs was comparable to the En4D--Var.
However, the estimates varied significantly based on the initial guess.
These ranged from $\SI{4.32}{kPa}$ to $\SI{9.46}{kPa}$ for shear modulus, and from $\SI{0.085}{Pa.s}$ to $\SI{0.114}{Pa.s}$ for viscosity.
Except for estimates from guess 1, the shear modulus estimates are also outside the bounds given by \citet{Estrada_2018}, and the radius fit is significantly worse than that of the En4D--Var, with an average NRMSE of $\num{9.10e-2}$.

On the other hand, while still worse than that of the En4D--Var, the IEnKS--MDA estimates are within the \citet{Estrada_2018} margin, and the radius fit is better than that of the IEnKS--SDA ($\mathrm{NRMSE}=\num{6.79e-2}$).
The IEnKS--MDA thus represents the best tested quasi-online method, as expected from the simulated data results of section~\ref{sec:results_sim}.
It is important to note here that the bubble radius fits were all obtained by re-running simulations with final shear modulus and viscosity estimates and comparing to experimental measurements.
This is a fair way to compare the ability of each method to estimate these parameters.
However, the radius fit obtained online during assimilation with the IEnKS methods is better (and comparable with the En4D--Var estimates), given that the radius is also being directly corrected at each time-step as part of the state vector.
For parameter estimation, the En4D--Var is the best tested method, but the IEnKS--MDA is a good quasi-online estimator.
This is particularly useful for the discussion in section~\ref{sec:exp_discussion}, where we make use of this time-varying estimation.

%%%%%

\subsection{Refined estimates} \label{sec:exp_discussion}

\begin{figure}[H]
    \centering
    \begin{subfigure}{.5\textwidth}
    \centering
    \includegraphics[width =\textwidth]{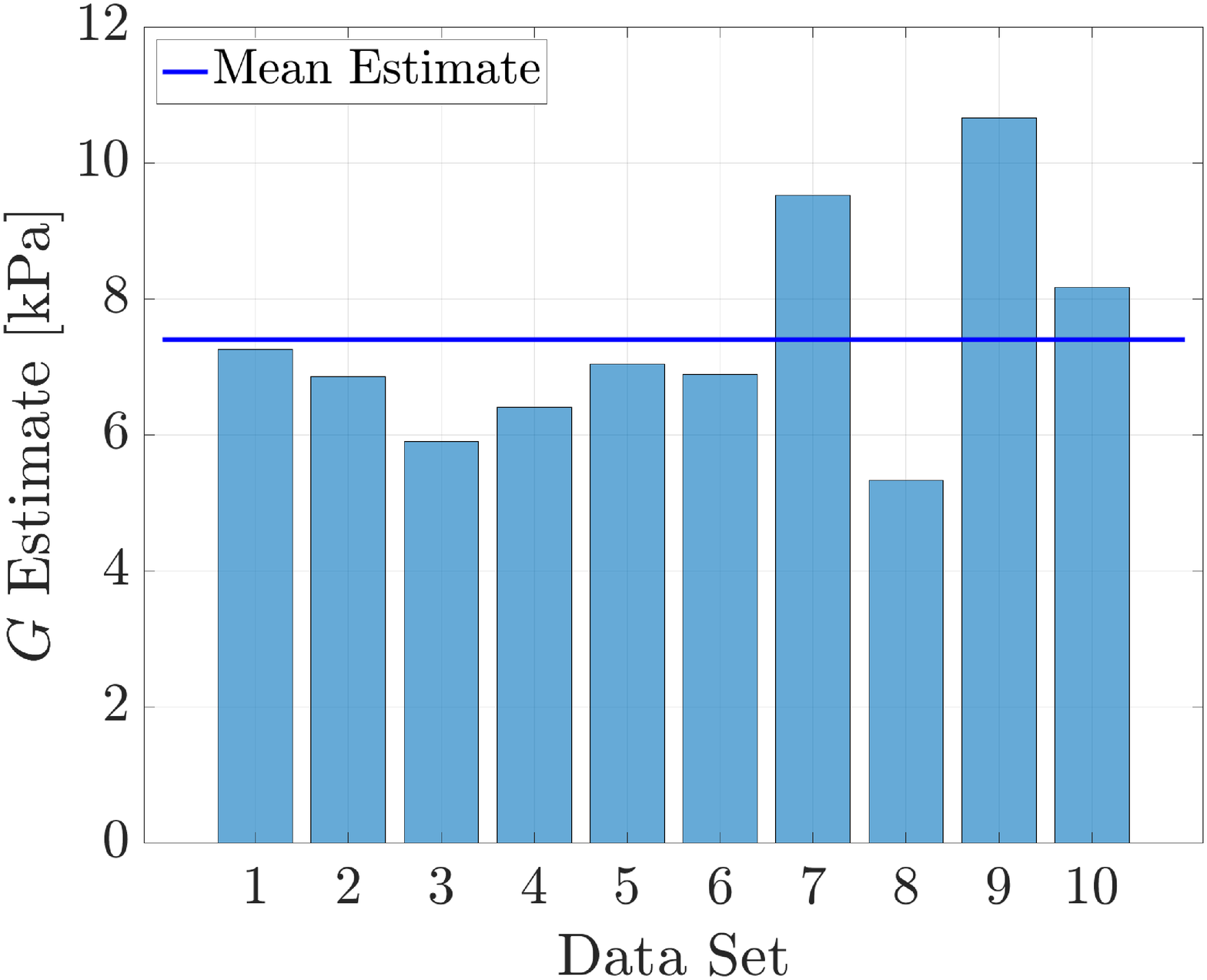}
    \caption{G estimates}
    \label{fig:exp_barplot_G}
    \end{subfigure}%
    \begin{subfigure}{.5\textwidth}
    \centering
    \includegraphics[width =\textwidth]{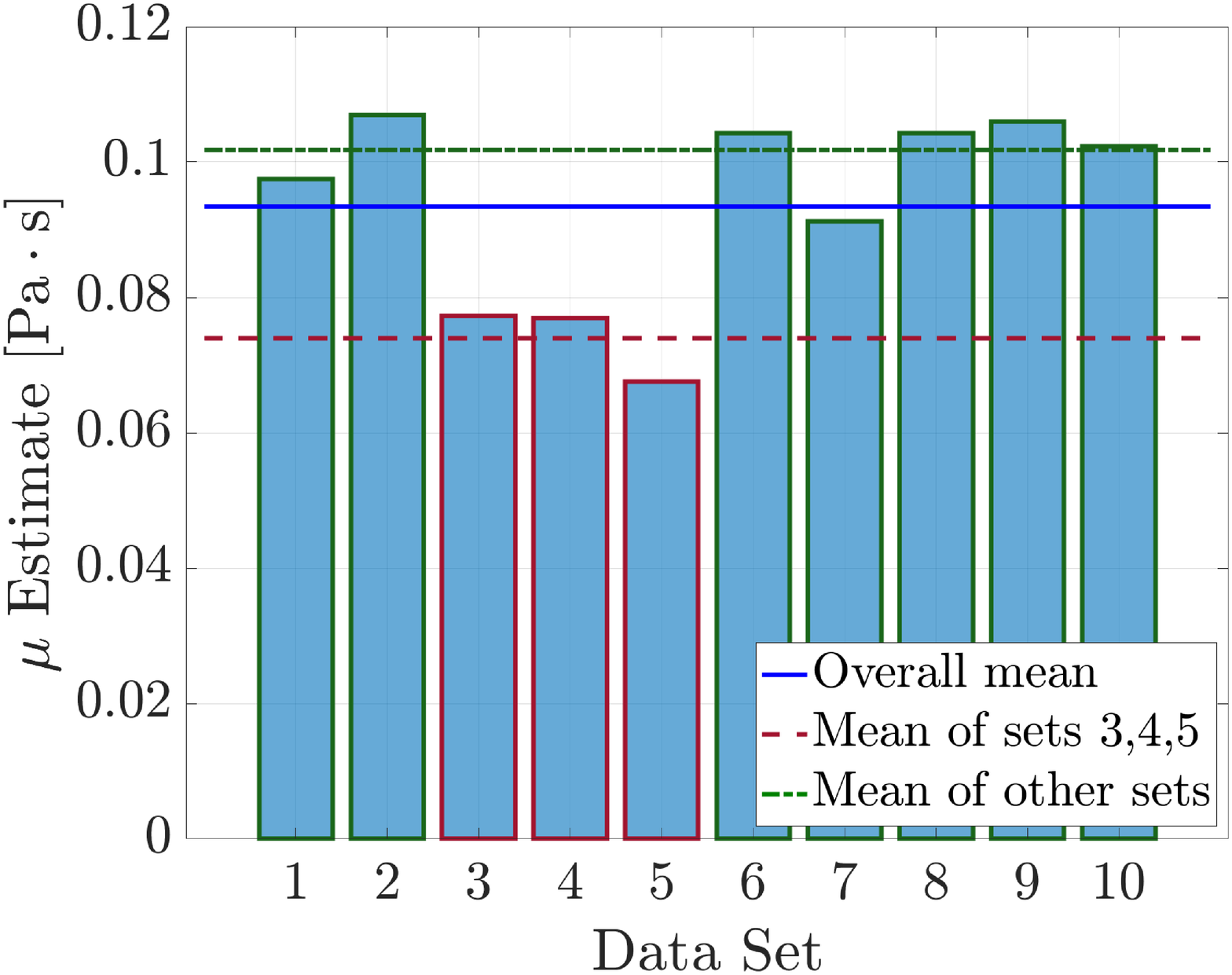}
    \caption{$\mu$ estimates}
    \label{fig:exp_barplot_mu}
    \end{subfigure}
    \caption{En4D--Var estimates for 10 experimental data sets.}
    \label{fig:exp_bar_plots}
\end{figure}

The estimates obtained for the shear modulus and viscosity from the previous section show that ensemble data assimilation methods can be effectively used for estimation of viscoelastic material properties.
A further look at the results, though, provides more information than simply this estimate.
Examining estimates for each variable across the 10 tested experimental data sets, as shown in figure~\ref{fig:exp_bar_plots}, there appears to be a discrepancy between data sets 3, 4, 5 and the rest for the viscosity.
While the shear modulus estimation shows no discernible trend (despite the previously mentioned larger spread in results), the viscosity data appears to be split between two estimates.
The red line in figure~\ref{fig:exp_barplot_mu} shows the mean estimate of data sets 3 to 5 ($\mu = \SI{0.074}{Pa.s}$) and the green line that of the rest ($\mu = \SI{0.102}{Pa.s}$).

\begin{figure}[H]
    \centering
    \begin{subfigure}{.5\textwidth}
    \centering
    \includegraphics[width =\textwidth]{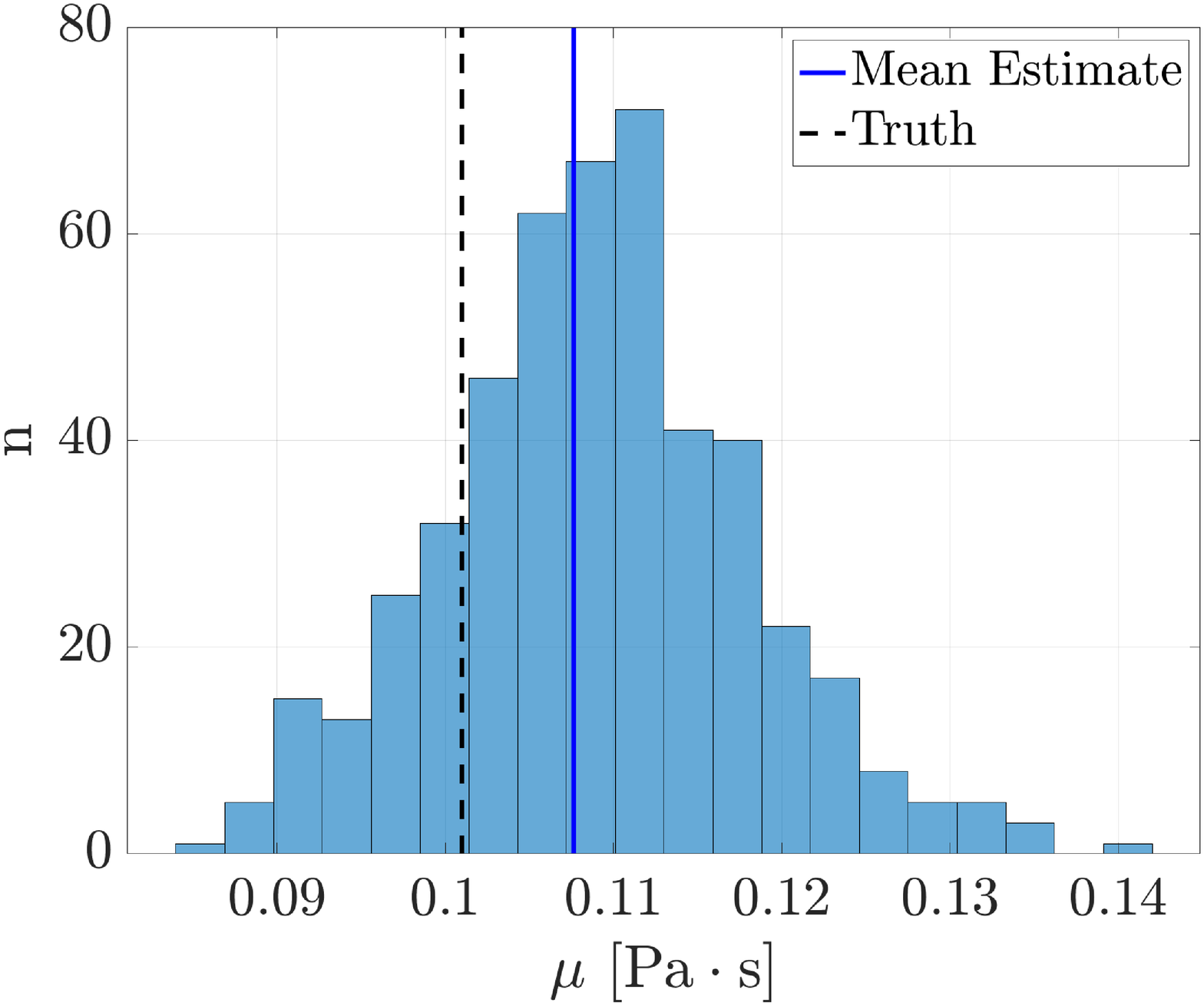}
    \caption{Simulated data}
    \label{fig:sim_hist_mu}
    \end{subfigure}%
    \begin{subfigure}{.5\textwidth}
    \centering
    \includegraphics[width =\textwidth]{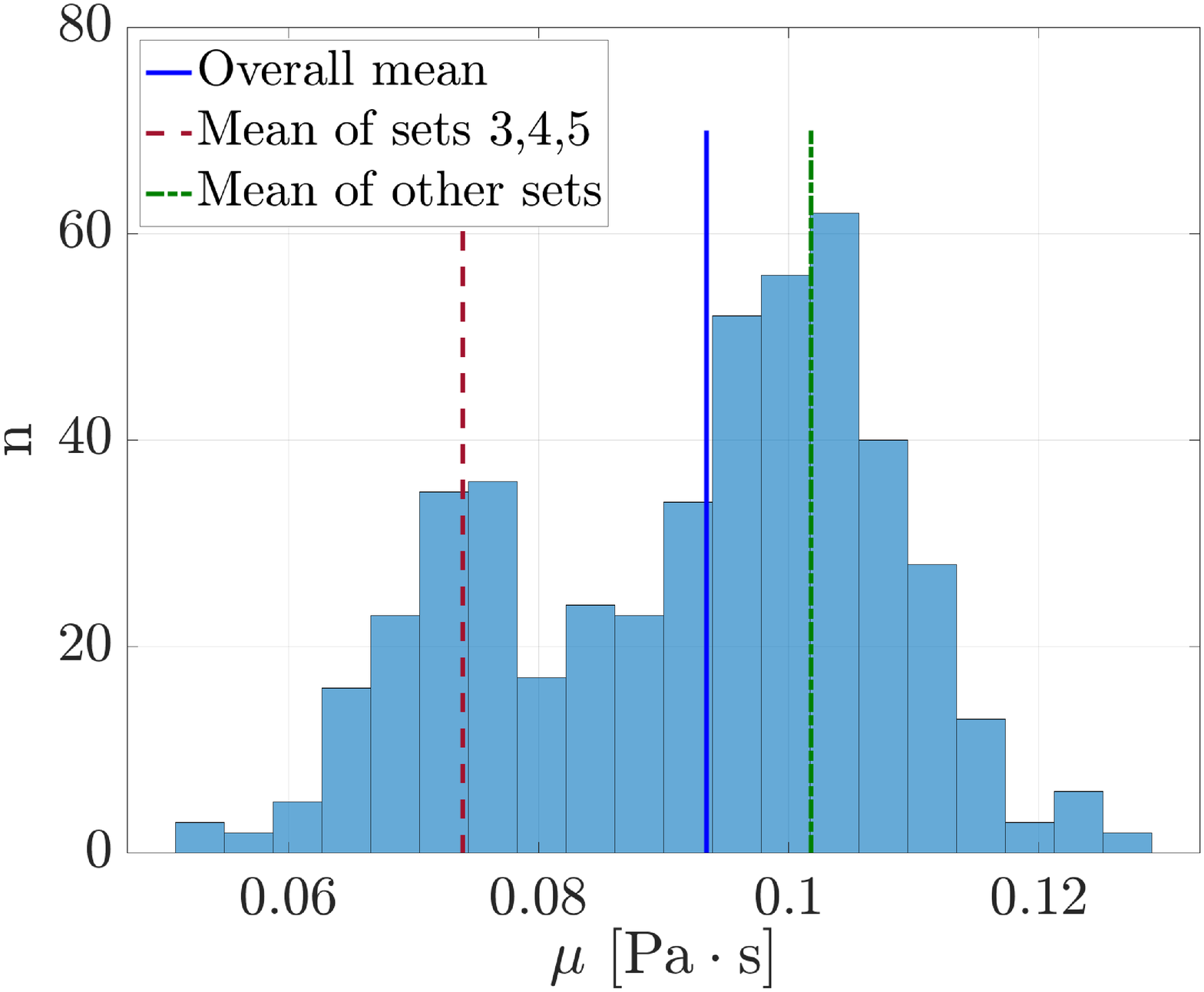}
    \caption{Experimental data}
    \label{fig:exp_hist_mu}
    \end{subfigure}
    \caption{Comparing final combined ensembles for viscosity estimation in simulated and experimental data.}
    \label{fig:hist_mu}
\end{figure}

Figure~\ref{fig:hist_mu} compares the histogram obtained when collating the $10 \times q$ final ensemble members for viscosity from 10 runs with different simulated data but the same ground truth (Figure~\ref{fig:sim_hist_mu}) and the 10 runs done with experimental data (Figure~\ref{fig:exp_hist_mu}).
As discussed in section~\ref{sec:sim_uncertainty}, we expect to approximately retrieve a Gaussian distribution around the estimate, as is the case for the simulated run in figure~\ref{fig:sim_hist_mu}.
However, figure~\ref{fig:exp_hist_mu} shows an apparent bimodal distribution.
The lower viscosity peak corresponds to the mean estimate of data sets 3 to 5, and the higher peak to that of the rest. \\

\begin{figure}[H]
    \centering
    \begin{subfigure}{.5\textwidth}
    \centering
    \includegraphics[width =\textwidth]{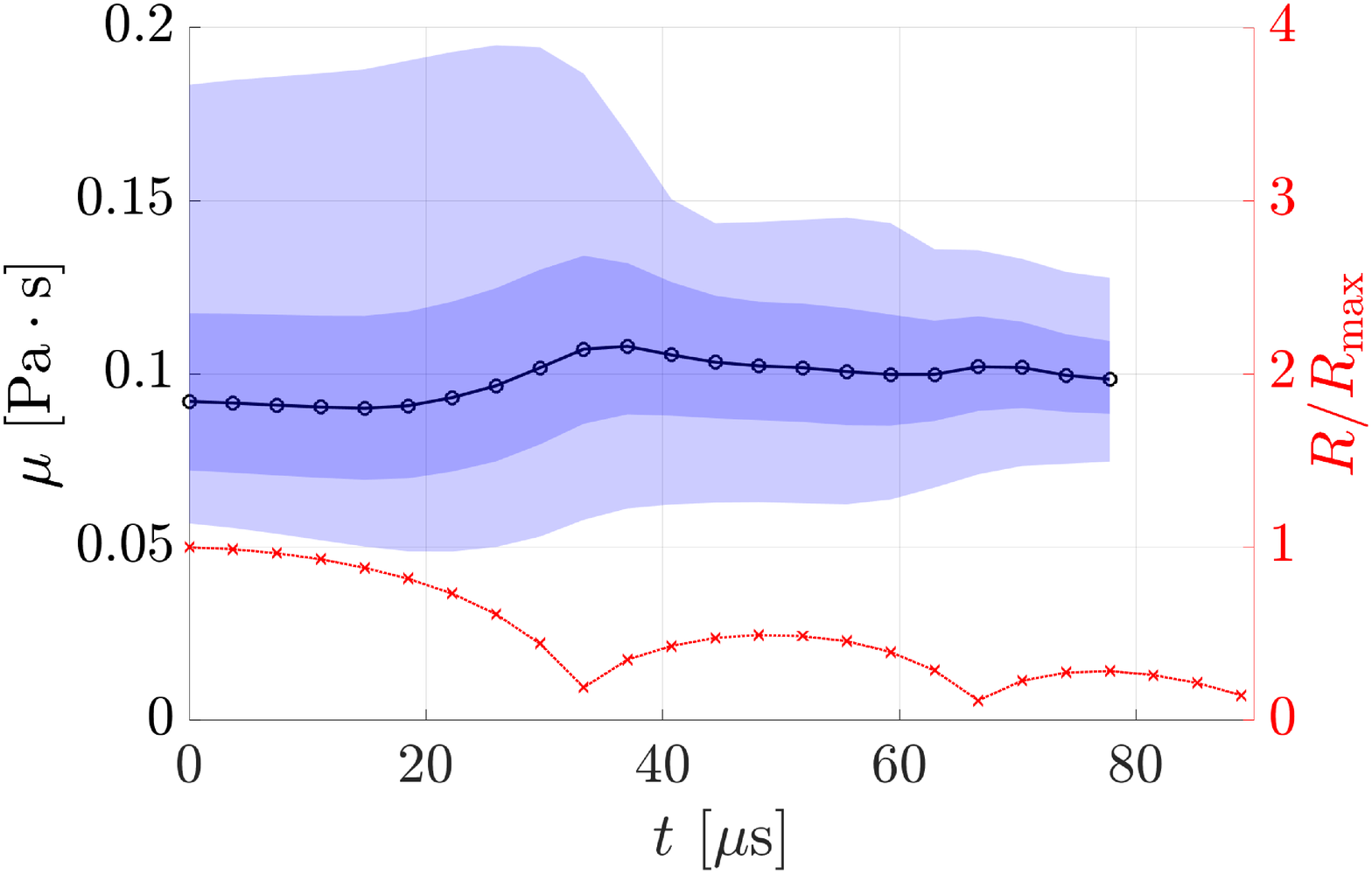}
    \caption{}
    \label{fig:exp_mda_set2}
    \end{subfigure}%
    \begin{subfigure}{.5\textwidth}
    \centering
    \includegraphics[width =\textwidth]{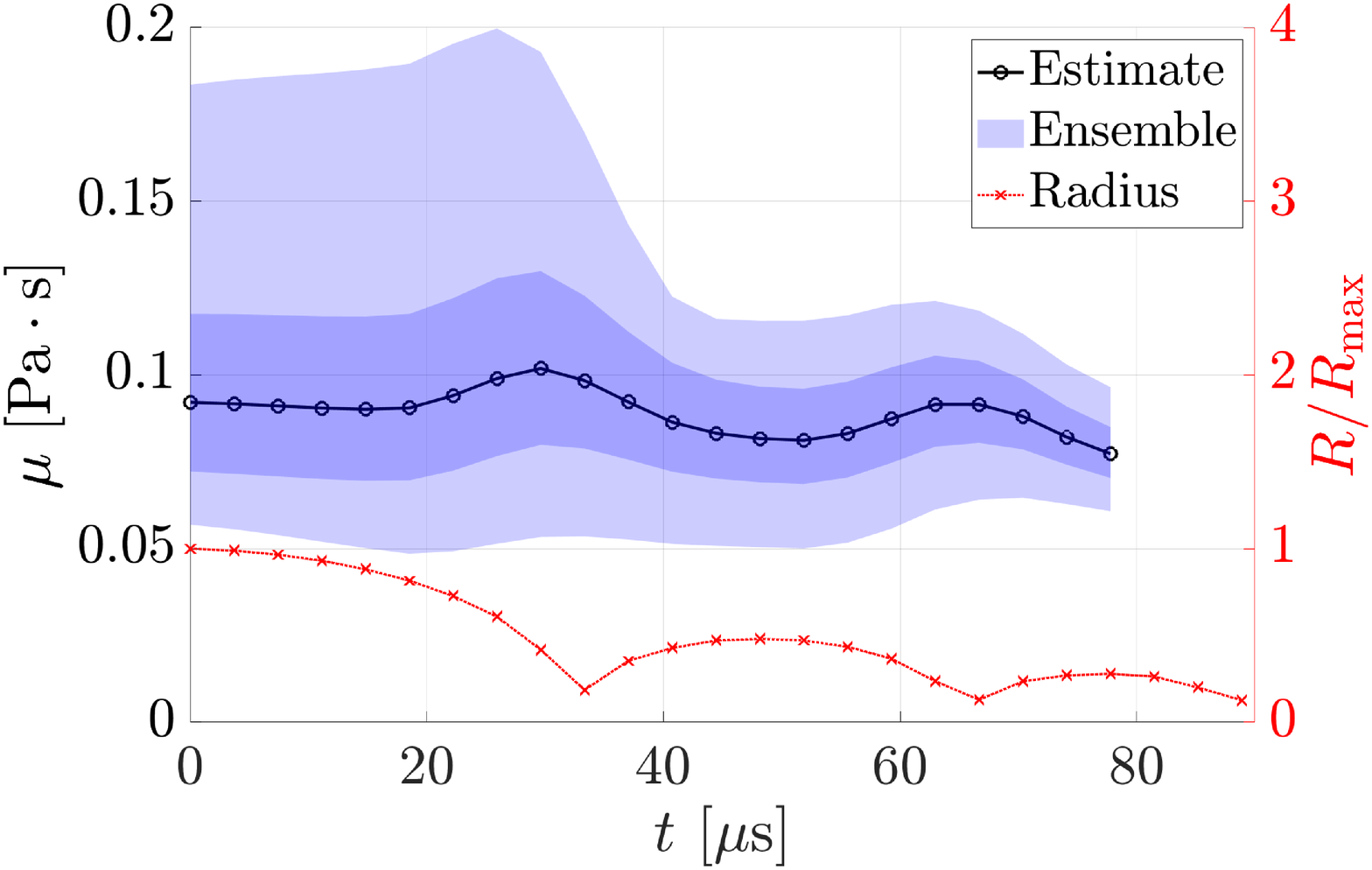}
    \caption{}
    \label{fig:exp_mda_set3}
    \end{subfigure}
    \caption{Comparing online estimation of viscosity in data sets 2 \textbf{(\subref{fig:exp_mda_set2})} and 3 \textbf{(\subref{fig:exp_mda_set3})} using the IEnKS--MDA.}
    \label{fig:exp_mda}
\end{figure}

To understand what may be causing this discrepancy of results, it is useful to consider the IEnKS--MDA and its quasi-online estimation of viscosity.
Figure~\ref{fig:exp_mda} shows a comparison between viscosity estimation for data set 2 (figure~\ref{fig:exp_mda_set2}) and data set 3 (figure~\ref{fig:exp_mda_set3}).
These are representative of data sets with a high and low viscosity estimate respectively.
They result in estimates of $\mu = \SI{0.098}{Pa.s}$ for data set 2, and $\mu = \SI{0.077}{Pa.s}$ for data set 3.
Comparing the two data sets, it appears that the assimilation begins similarly, correcting to a higher viscosity estimate during the first collapse.
However, there is a divergence between the behavior of the smoother after each collapse point, particularly the second one (around $t=\SI{65}{\micro\second}$).
Figure~\ref{fig:exp_mda_set2} shows a slow decrease and convergence towards a higher viscosity value, with negligible change at the second collapse point.
However, this estimate drops sharply after these collapse points in figure~\ref{fig:exp_mda_set3}.
In fact, the data around each collapse causes the viscosity estimate to sharply drop in data set 3, which does not occur in data set 2.
This behavior is representative of what is seen in data sets 3 through 5, but does not occur in the rest of the runs.

Given the physical model used, the viscosity should be constant and such drops in the parameter are not expected.
The model alone thus cannot adequately capture the behavior of the gel seen by the IEnKS in these data sets.
We can posit that a violent collapse in these data sets is causing inelastic behavior in the material, and thereupon this perceived change in material properties \citep{Yang_2020}.
More work will be needed to determine the exact cause, but this could perhaps result from fracture, damage to the polymer network in the gel, or combustion in the gas phase \citep{Movahed_2016, Kundu_2009, Raayai_2019}.
Regardless of physical cause, this time-dependent behavior is not accounted for in the model, but is captured by the IEnKS--MDA as a drop in the perceived viscosity.

\begin{figure}[H]
    \centering
    \includegraphics[width=.5\textwidth]{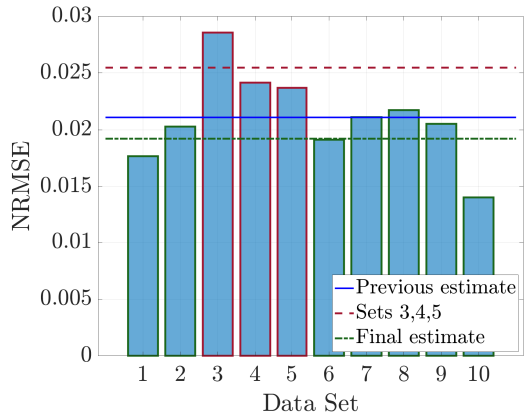}
    \caption{Bar plot of radius normalized root mean squared errors for each data set. Also plotted are the previous estimate mean NRMSE (mean of all sets), the mean NRMSE for sets 3 to 5, and the final estimate mean NRMSE (mean of all other sets).}
    \label{fig:exp_En4D_NRMSE}
\end{figure}

Figure~\ref{fig:exp_En4D_NRMSE} compares the normalized root mean squared error (NRMSE) across all data sets, given by
\begin{gather} \label{eq:nrmse}
    \mathrm{NRMSE} = \frac{\sqrt{\overline{\left(\bm{y}_{\mathrm{sim}} - \bm{y}_{\mathrm{exp}}\right)^2}}}{\overline{\bm{y}_{\mathrm{exp}}}},
\end{gather}
where $\bm{y}_{\mathrm{exp}}$ is the experimental bubble radius time history, and $\bm{y}_{\mathrm{sim}}$ the simulated time history given the estimated material properties (at the corresponding times).

Figure~\ref{fig:exp_En4D_NRMSE} shows a higher error in the estimated bubble radius curves fit for data sets 3 through 5, which is expected given the heightened model uncertainty in these data sets.
Because of this uncertainty and higher error, we discard these three sets as outliers, which yields the final IEnKS--MDA-informed En4D--Var estimate reported in table \ref{tab:exp_final_results}.
Notable are the drop in standard deviation for viscosity as compared to the previous En4D--Var estimate and the reduced NRMSE.

\begin{table}[H]
    \centering
    \begin{tabular}{|c|c|c|c|}
\hline
    \textbf{Estimate} & \textbf{$\bm{G \pm \sigma}$ [\si{kPa}]} & \textbf{$\bm{\mu \pm \sigma}$ [\si{Pa.s}]} & \textbf{NRMSE} \\
\hline
    \citet{Estrada_2018} & $7.69 \pm 1.12$ & $0.101 \pm 0.023$& \\
\hline
    Previous & $7.41 \pm 1.63$ & $0.093 \pm 0.014$ & $\num{2.16e-2}$ \\
%\hline
    \textbf{Final} & $\mathbf{7.81 \pm 1.80}$ & $\mathbf{0.102 \pm 0.006 }$ & $\mathbf{\num{1.95e-2}}$ \\
\hline
    \end{tabular}
    \caption{Final En4D--Var estimates (discarding three outlier data sets) and standard deviation, along with the average radius normalized root mean squared error. The previous best estimate corresponds to the mean of all 10 data sets, outliers included.}
    \label{tab:exp_final_results}
\end{table}

%%%%%%%%%%%%%%%%%%%%%%%%%%%%%%
%%%%%%%% Conclusions %%%%%%%%%
%%%%%%%%%%%%%%%%%%%%%%%%%%%%%%

\section{Conclusions} \label{sec:conclusion}

Ensemble-based data assimilation was successfully used to estimate the mechanical properties of soft viscoelastic materials at high strain rates via observations of bubble collapse.
In particular, the ensemble-based 4D--var method (En4D--Var) provided an accurate estimate efficiently, while the iterative ensemble Kalman smoother with multiple data assimilation (IEnKS--MDA) reliably estimated parameters quasi-online.
Added benefits of these algorithms include adaptability to different numerical or viscoelastic models, and scalability to further parameter estimation, with negligible computational cost for additional parameters.
These methods account for both model and experimental error, with noisy measurements or inaccuracies in the model having a limited impact on estimation.
The ability to adjust the entire state vector rather than just the parameters to estimate limits inaccuracies from a poor initial guess.
Overall, this represents a viable framework for estimation of mechanical properties of viscoelastic materials.

Using the En4D--Var and IEnKS--MDA together provided information, in this case, in the form of model error.
It is hypothesized that the bubble collapses are damaging the polyacrylamide gel in certain test cases, leading to a reduced estimated viscosity after each subsequent collapse, a physical effect not accounted for in the model.
Thus, while the En4D--Var provides the best estimates (especially given its relative computational efficiency), the IEnKS--MDA can provide additional information about time-dependent modeling errors.

%%%%%%%%%%%%%%%%%%%%%%%%%%%%%%
%%%%% Acknowledgements %%%%%%%
%%%%%%%%%%%%%%%%%%%%%%%%%%%%%%

\section*{Acknowledgements}

This work was supported by the National Institutes of Health [grant number 2P01-DK043881]; and the Office of Naval Research [grant numbers N0014-18-1-2625, N0014-17-1-2676].

\bibliography{bibliography}

\begin{thebibliography}{45}
\providecommand{\natexlab}[1]{#1}
\providecommand{\url}[1]{\texttt{#1}}
\expandafter\ifx\csname urlstyle\endcsname\relax
  \providecommand{\doi}[1]{doi: #1}\else
  \providecommand{\doi}{doi: \begingroup \urlstyle{rm}\Url}\fi

\bibitem[Akhatov et~al.(2001)Akhatov, Lindau, Topolnikov, Mettin, Vakhitova,
  and Lauterborn]{Akhatov_2001}
I.~Akhatov, O.~Lindau, A.~Topolnikov, R.~Mettin, N.~Vakhitova, and
  W.~Lauterborn.
\newblock Collapse and rebound of a laser-induced cavitation bubble.
\newblock \emph{Physics of Fluids}, 13\penalty0 (10):\penalty0 2805--2819,
  2001.
\newblock \doi{10.1063/1.1401810}.
\newblock URL \url{https://doi.org/10.1063/1.1401810}.

\bibitem[Bailey et~al.(2003)Bailey, Khokhlova, Sapozhnikov, Kargl, and
  Crum]{Bailey_2003}
M.~Bailey, V.~Khokhlova, O.~Sapozhnikov, S.~Kargl, and L.~Crum.
\newblock Physical mechanisms of the therapeutic effect of ultrasound (a
  review).
\newblock \emph{Acoust Phys}, 49:\penalty0 369--388, 07 2003.
\newblock \doi{10.1134/1.1591291}.
\newblock URL \url{http://dx.doi.org/10.1134/1.1591291}.

\bibitem[Bar-Kochba et~al.(2016)Bar-Kochba, Scimone, Estrada, and
  Franck]{Bar-Kochba_2016}
E.~Bar-Kochba, M.~T. Scimone, J.~B. Estrada, and C.~Franck.
\newblock Strain and rate-dependent neuronal injury in a 3d in vitro
  compression model of traumatic brain injury.
\newblock \emph{Scientific Reports}, 6\penalty0 (1):\penalty0 30550, 2016.
\newblock \doi{10.1038/srep30550}.
\newblock URL \url{https://doi.org/10.1038/srep30550}.

\bibitem[Barajas and Johnsen(2017)]{Barajas_2017}
C.~Barajas and E.~Johnsen.
\newblock The effects of heat and mass diffusion on freely oscillating bubbles
  in a viscoelastic, tissue-like medium.
\newblock \emph{The Journal of the Acoustical Society of America}, 141\penalty0
  (2):\penalty0 908--918, Feb 2017.
\newblock ISSN 0001-4966.
\newblock \doi{10.1121/1.4976081}.
\newblock URL \url{http://dx.doi.org/10.1121/1.4976081}.

\bibitem[Bocquet(2011)]{Bocquet_2011}
M.~Bocquet.
\newblock Ensemble {K}alman filtering without the intrinsic need for inflation.
\newblock \emph{Nonlinear Processes in Geophysics}, 18\penalty0 (5):\penalty0
  735--750, 2011.
\newblock \doi{10.5194/npg-18-735-2011}.
\newblock URL \url{https://www.nonlin-processes-geophys.net/18/735/2011/}.

\bibitem[Bocquet and Sakov(2013{\natexlab{a}})]{Bocquet_2013}
M.~Bocquet and P.~Sakov.
\newblock An iterative ensemble {K}alman smoother.
\newblock \emph{Quarterly Journal of the Royal Meteorological Society},
  140\penalty0 (682):\penalty0 1521--1535, Oct 2013{\natexlab{a}}.
\newblock \doi{10.1002/qj.2236}.
\newblock URL \url{http://dx.doi.org/10.1002/qj.2236}.

\bibitem[Bocquet and Sakov(2013{\natexlab{b}})]{Bocquet_2013_2}
M.~Bocquet and P.~Sakov.
\newblock Joint state and parameter estimation with an iterative ensemble
  {K}alman smoother.
\newblock \emph{Nonlinear Processes in Geophysics}, 20\penalty0 (5):\penalty0
  803--818, Oct 2013{\natexlab{b}}.
\newblock \doi{10.5194/npg-20-803-2013}.
\newblock URL \url{http://dx.doi.org/10.5194/npg-20-803-2013}.

\bibitem[Caya et~al.(2005)Caya, Sun, and Snyder]{Caya_2005}
A.~Caya, J.~Sun, and C.~Snyder.
\newblock A comparison between the {4DVAR} and the ensemble {K}alman filter
  techniques for radar data assimilation.
\newblock \emph{Monthly Weather Review}, 133\penalty0 (11):\penalty0
  3081--3094, 2005.
\newblock \doi{10.1175/MWR3021.1}.
\newblock URL \url{https://doi.org/10.1175/MWR3021.1}.

\bibitem[Epstein and Keller(1972)]{Epstein_1972}
D.~Epstein and J.~B. Keller.
\newblock Expansion and contraction of planar, cylindrical, and spherical
  underwater gas bubbles.
\newblock \emph{The Journal of the Acoustical Society of America}, 52\penalty0
  (3B):\penalty0 975--980, 1972.
\newblock \doi{10.1121/1.1913203}.
\newblock URL \url{https://doi.org/10.1121/1.1913203}.

\bibitem[Estrada et~al.(2018)Estrada, Barajas, Henann, Johnsen, and
  Franck]{Estrada_2018}
J.~B. Estrada, C.~Barajas, D.~L. Henann, E.~Johnsen, and C.~Franck.
\newblock High strain-rate soft material characterization via inertial
  cavitation.
\newblock \emph{Journal of the Mechanics and Physics of Solids}, 112:\penalty0
  291--317, 2018.
\newblock \doi{https://doi.org/10.1016/j.jmps.2017.12.006}.
\newblock URL
  \url{http://www.sciencedirect.com/science/article/pii/S0022509617307585}.

\bibitem[Evensen(1994)]{Evensen_1994}
G.~Evensen.
\newblock Sequential data assimilation with a nonlinear quasi-geostrophic model
  using {M}onte {C}arlo methods to forecast error statistics.
\newblock \emph{Journal of Geophysical Research: Oceans}, 99\penalty0
  (C5):\penalty0 10143--10162, 1994.
\newblock \doi{10.1029/94JC00572}.
\newblock URL
  \url{https://agupubs.onlinelibrary.wiley.com/doi/abs/10.1029/94JC00572}.

\bibitem[Evensen(2003)]{Evensen_2003}
G.~Evensen.
\newblock The ensemble {K}alman filter: {T}heoretical formulation and practical
  implementation.
\newblock \emph{Ocean Dynamics}, 53\penalty0 (4):\penalty0 343--367, Nov 2003.
\newblock \doi{10.1007/s10236-003-0036-9}.
\newblock URL \url{http://dx.doi.org/10.1007/s10236-003-0036-9}.

\bibitem[Evensen(2009{\natexlab{a}})]{Evensen_2009}
G.~Evensen.
\newblock \emph{Data Assimilation: {T}he Ensemble {K}alman Filter}.
\newblock Springer, 2009{\natexlab{a}}.

\bibitem[Evensen(2009{\natexlab{b}})]{Evensen_2009_2}
G.~Evensen.
\newblock The ensemble {K}alman filter for combined state and parameter
  estimation.
\newblock \emph{IEEE Control Systems Magazine}, 29\penalty0 (3):\penalty0
  83--104, June 2009{\natexlab{b}}.
\newblock \doi{10.1109/MCS.2009.932223}.
\newblock URL \url{http://dx.doi.org/10.1109/MCS.2009.932223}.

\bibitem[Evensen and van Leeuwen(2000)]{Evensen_2000}
G.~Evensen and P.~J. van Leeuwen.
\newblock An ensemble {K}alman smoother for nonlinear dynamics.
\newblock \emph{Monthly Weather Review}, 128:\penalty0 1852--1867, Jun 2000.

\bibitem[Gaudron et~al.(2015)Gaudron, Warnez, and Johnsen]{Gaudron_2015}
R.~Gaudron, M.~T. Warnez, and E.~Johnsen.
\newblock Bubble dynamics in a viscoelastic medium with nonlinear elasticity.
\newblock \emph{Journal of Fluid Mechanics}, 766:\penalty0 54--75, Jan 2015.
\newblock \doi{10.1017/jfm.2015.7}.
\newblock URL \url{http://dx.doi.org/10.1017/jfm.2015.7}.

\bibitem[Gustafsson and Bojarova(2014)]{Gustafsson_2014}
N.~Gustafsson and J.~Bojarova.
\newblock Four-dimensional ensemble variational (4d-en-var) data assimilation
  for the high resolution limited area model (hirlam).
\newblock \emph{Nonlinear processes in geophysics}, 21\penalty0 (4):\penalty0
  745--762, 2014.

\bibitem[Hosea and Shampine(1996)]{Hosea_1996}
M.~Hosea and L.~Shampine.
\newblock Analysis and implementation of {TR--BDF2}.
\newblock \emph{Applied Numerical Mathematics}, 20\penalty0 (1):\penalty0
  21--37, 1996.
\newblock \doi{https://doi.org/10.1016/0168-9274(95)00115-8}.
\newblock URL
  \url{http://www.sciencedirect.com/science/article/pii/0168927495001158}.

\bibitem[Kalman(1960)]{Kalman_1960}
R.~E. Kalman.
\newblock A new approach to linear filtering and prediction problems.
\newblock \emph{Journal of Basic Engineering}, 82\penalty0 (1):\penalty0
  35--45, 1960.

\bibitem[Katzfuss et~al.(2016)Katzfuss, Stroud, and Wikle]{Katzfuss_2016}
M.~Katzfuss, J.~R. Stroud, and C.~K. Wikle.
\newblock Understanding the ensemble {K}alman filter.
\newblock \emph{The American Statistician}, 70\penalty0 (4):\penalty0 350--357,
  2016.
\newblock \doi{10.1080/00031305.2016.1141709}.
\newblock URL \url{https://doi.org/10.1080/00031305.2016.1141709}.

\bibitem[Keller and Miksis(1980)]{Keller_1980}
J.~B. Keller and M.~Miksis.
\newblock Bubble oscillations of large amplitude.
\newblock \emph{The Journal of the Acoustical Society of America}, 68\penalty0
  (2):\penalty0 628--633, Aug 1980.
\newblock ISSN 0001-4966.
\newblock \doi{10.1121/1.384720}.
\newblock URL \url{http://dx.doi.org/10.1121/1.384720}.

\bibitem[Kundu and Crosby(2009)]{Kundu_2009}
S.~Kundu and A.~J. Crosby.
\newblock Cavitation and fracture behavior of polyacrylamide hydrogels.
\newblock \emph{Soft Matter}, 5:\penalty0 3963--3968, 2009.
\newblock \doi{10.1039/B909237D}.
\newblock URL \url{http://dx.doi.org/10.1039/B909237D}.

\bibitem[Liu et~al.(2008)Liu, Xiao, and Wang]{Liu_2008}
C.~Liu, Q.~Xiao, and B.~Wang.
\newblock {An Ensemble-Based Four-Dimensional Variational Data Assimilation
  Scheme. Part I: Technical Formulation and Preliminary Test}.
\newblock \emph{Monthly Weather Review}, 136\penalty0 (9):\penalty0 3363--3373,
  09 2008.

\bibitem[Luo and Hoteit(2011)]{Luo_2011}
X.~Luo and I.~Hoteit.
\newblock Robust ensemble filtering and its relation to covariance inflation in
  the ensemble {K}alman filter.
\newblock \emph{Monthly Weather Review}, 139\penalty0 (12):\penalty0
  3938--3953, 2011.
\newblock \doi{10.1175/MWR-D-10-05068.1}.
\newblock URL \url{https://doi.org/10.1175/MWR-D-10-05068.1}.

\bibitem[Mancia et~al.(2017)Mancia, Vlaisavljevich, Xu, and
  Johnsen]{Mancia_2017}
L.~Mancia, E.~Vlaisavljevich, Z.~Xu, and E.~Johnsen.
\newblock Predicting tissue susceptibility to mechanical cavitation damage in
  therapeutic ultrasound.
\newblock \emph{Ultrasound in Medicine \& Biology}, 43\penalty0 (7):\penalty0
  1421 -- 1440, 2017.
\newblock \doi{https://doi.org/10.1016/j.ultrasmedbio.2017.02.020}.
\newblock URL
  \url{http://www.sciencedirect.com/science/article/pii/S030156291730100X}.

\bibitem[Maxwell et~al.(2009)Maxwell, Cain, Duryea, Yuan, Gurm, and
  Xu]{Maxwell_2009}
A.~D. Maxwell, C.~A. Cain, A.~P. Duryea, L.~Yuan, H.~S. Gurm, and Z.~Xu.
\newblock Noninvasive thrombolysis using pulsed ultrasound cavitation therapy
  – histotripsy.
\newblock \emph{Ultrasound in Medicine \& Biology}, 35\penalty0 (12):\penalty0
  1982 -- 1994, 2009.
\newblock \doi{https://doi.org/10.1016/j.ultrasmedbio.2009.07.001}.
\newblock URL
  \url{http://www.sciencedirect.com/science/article/pii/S0301562909014252}.

\bibitem[Meaney and Smith(2011)]{Meaney_2011}
D.~F. Meaney and D.~H. Smith.
\newblock Biomechanics of concussion.
\newblock \emph{Clinics in Sports Medicine}, 30\penalty0 (1):\penalty0 19 --
  31, 2011.
\newblock \doi{https://doi.org/10.1016/j.csm.2010.08.009}.
\newblock URL
  \url{http://www.sciencedirect.com/science/article/pii/S027859191000061X}.
\newblock Concussion in Sports.

\bibitem[Movahed et~al.(2016)Movahed, Kreider, Maxwell, Hutchens, and
  Freund]{Movahed_2016}
P.~Movahed, W.~Kreider, A.~D. Maxwell, S.~B. Hutchens, and J.~B. Freund.
\newblock Cavitation-induced damage of soft materials by focused ultrasound
  bursts: A fracture-based bubble dynamics model.
\newblock \emph{The Journal of the Acoustical Society of America}, 140\penalty0
  (2):\penalty0 1374--1386, 2016.
\newblock \doi{10.1121/1.4961364}.
\newblock URL \url{https://doi.org/10.1121/1.4961364}.

\bibitem[Nyein et~al.(2010)Nyein, Jason, Yu, Pita, Joannopoulos, Moore, and
  Radovitzky]{Nyein_2010}
M.~K. Nyein, A.~M. Jason, L.~Yu, C.~M. Pita, J.~D. Joannopoulos, D.~F. Moore,
  and R.~A. Radovitzky.
\newblock In silico investigation of intracranial blast mitigation with
  relevance to military traumatic brain injury.
\newblock \emph{Proceedings of the National Academy of Sciences}, 107\penalty0
  (48):\penalty0 20703--20708, 2010.
\newblock \doi{10.1073/pnas.1014786107}.
\newblock URL \url{https://www.pnas.org/content/107/48/20703}.

\bibitem[Preston et~al.(2007)Preston, Colonius, and Brennen]{Preston_2007}
A.~T. Preston, T.~Colonius, and C.~E. Brennen.
\newblock A reduced-order model of diffusive effects on the dynamics of
  bubbles.
\newblock \emph{Physics of Fluids}, 19\penalty0 (12):\penalty0 123302, 2007.
\newblock \doi{10.1063/1.2825018}.
\newblock URL \url{https://doi.org/10.1063/1.2825018}.

\bibitem[Prosperetti and Lezzi(1986)]{Prosperetti_1986}
A.~Prosperetti and A.~Lezzi.
\newblock Bubble dynamics in a compressible liquid. part 1. first-order theory.
\newblock \emph{Journal of Fluid Mechanics}, 168:\penalty0 457–478, 1986.
\newblock \doi{10.1017/S0022112086000460}.
\newblock URL \url{http://dx.doi.org/10.1017/S0022112086000460}.

\bibitem[Prosperetti et~al.(1988)Prosperetti, Crum, and
  Commander]{Prosperetti_1988}
A.~Prosperetti, L.~A. Crum, and K.~W. Commander.
\newblock Nonlinear bubble dynamics.
\newblock \emph{The Journal of the Acoustical Society of America}, 83\penalty0
  (2):\penalty0 502--514, 1988.
\newblock \doi{10.1121/1.396145}.
\newblock URL \url{https://doi.org/10.1121/1.396145}.

\bibitem[Raayai-Ardakani et~al.(2019)Raayai-Ardakani, Earl, and
  Cohen]{Raayai_2019}
S.~Raayai-Ardakani, D.~R. Earl, and T.~Cohen.
\newblock The intimate relationship between cavitation and fracture.
\newblock \emph{Soft Matter}, 15:\penalty0 4999--5005, 2019.
\newblock \doi{10.1039/C9SM00570F}.
\newblock URL \url{http://dx.doi.org/10.1039/C9SM00570F}.

\bibitem[Sakov et~al.(2012)Sakov, Oliver, and Bertino]{Sakov_2012}
P.~Sakov, D.~S. Oliver, and L.~Bertino.
\newblock An iterative {EnKF} for strongly nonlinear systems.
\newblock \emph{Monthly Weather Review}, 140\penalty0 (6):\penalty0 1988--2004,
  Jun 2012.
\newblock \doi{10.1175/mwr-d-11-00176.1}.
\newblock URL \url{http://dx.doi.org/10.1175/MWR-D-11-00176.1}.

\bibitem[Sarntinoranont et~al.(2012)Sarntinoranont, Lee, Hong, King, Subhash,
  Kwon, and Moore]{Sartinoranont_2012}
M.~Sarntinoranont, S.~J. Lee, Y.~Hong, M.~A. King, G.~Subhash, J.~Kwon, and
  D.~F. Moore.
\newblock High-strain-rate brain injury model using submerged acute rat brain
  tissue slices.
\newblock \emph{Journal of Neurotrauma}, 29\penalty0 (2):\penalty0 418--429,
  2012.
\newblock \doi{10.1089/neu.2011.1772}.
\newblock URL \url{https://doi.org/10.1089/neu.2011.1772}.
\newblock PMID: 21970544.

\bibitem[Schillings and Stuart(2017)]{Schillings_2017}
C.~Schillings and A.~M. Stuart.
\newblock Analysis of the ensemble {K}alman filter for inverse problems.
\newblock \emph{SIAM Journal on Numerical Analysis}, 55\penalty0 (3):\penalty0
  1264--1290, 2017.

\bibitem[Tr{\'e}molet(2007)]{Tremolet_2007}
Y.~Tr{\'e}molet.
\newblock Incremental {4D--Var} convergence study.
\newblock \emph{Tellus A: Dynamic Meteorology and Oceanography}, 59\penalty0
  (5):\penalty0 706--718, 2007.
\newblock \doi{10.1111/j.1600-0870.2007.00271.x}.
\newblock URL \url{https://doi.org/10.1111/j.1600-0870.2007.00271.x}.

\bibitem[van Leeuwen(1999)]{Leeuwen_1999}
P.~J. van Leeuwen.
\newblock Comment on ``{D}ata assimilation using an ensemble {K}alman filter
  technique''.
\newblock \emph{Monthly Weather Review}, 127\penalty0 (6):\penalty0 1374--1377,
  1999.
\newblock \doi{10.1175/1520-0493(1999)127<1374:CODAUA>2.0.CO;2}.
\newblock URL
  \url{https://doi.org/10.1175/1520-0493(1999)127<1374:CODAUA>2.0.CO;2}.

\bibitem[Vlaisavljevich et~al.(2015)Vlaisavljevich, Lin, Warnez, Singh, Mancia,
  Putnam, Johnsen, Cain, and Xu]{Vlaisavljevich_2015}
E.~Vlaisavljevich, K.-W. Lin, M.~T. Warnez, R.~Singh, L.~Mancia, A.~J. Putnam,
  E.~Johnsen, C.~Cain, and Z.~Xu.
\newblock Effects of tissue stiffness, ultrasound frequency, and pressure on
  histotripsy-induced cavitation bubble behavior.
\newblock \emph{Physics in medicine and biology}, 60\penalty0 (6):\penalty0
  2271--2292, 2015.
\newblock \doi{10.1088/0031-9155/60/6/2271}.
\newblock URL \url{http://dx.doi.org/10.1088/0031-9155/60/6/2271}.

\bibitem[Warnez and Johnsen(2015)]{Warnez_2015}
M.~T. Warnez and E.~Johnsen.
\newblock Numerical modeling of bubble dynamics in viscoelastic media with
  relaxation.
\newblock \emph{Physics of Fluids}, 27\penalty0 (6):\penalty0 063103, Jun 2015.
\newblock \doi{10.1063/1.4922598}.
\newblock URL \url{http://dx.doi.org/10.1063/1.4922598}.

\bibitem[Whitaker and Hamill(2012)]{Whitaker_2012}
J.~S. Whitaker and T.~M. Hamill.
\newblock Evaluating methods to account for system errors in ensemble data
  assimilation.
\newblock \emph{Monthly Weather Review}, 140\penalty0 (9):\penalty0 3078--3089,
  2012.
\newblock \doi{10.1175/MWR-D-11-00276.1}.
\newblock URL \url{https://doi.org/10.1175/MWR-D-11-00276.1}.

\bibitem[{Xu} et~al.(2007){Xu}, {Raghavan}, {Hall}, {Chang}, {Mycek},
  {Fowlkes}, and {Cain}]{Xu_2007}
Z.~{Xu}, M.~{Raghavan}, T.~L. {Hall}, C.~{Chang}, M.~{Mycek}, J.~B. {Fowlkes},
  and C.~A. {Cain}.
\newblock High speed imaging of bubble clouds generated in pulsed ultrasound
  cavitational therapy - histotripsy.
\newblock \emph{IEEE Transactions on Ultrasonics, Ferroelectrics, and Frequency
  Control}, 54\penalty0 (10):\penalty0 2091--2101, 2007.

\bibitem[Yang et~al.(2020)Yang, Cramer, and Franck]{Yang_2020}
J.~Yang, H.~C. Cramer, and C.~Franck.
\newblock Extracting non-linear viscoelastic material properties from
  violently-collapsing cavitation bubbles.
\newblock \emph{Extreme Mechanics Letters}, 39:\penalty0 100839, 2020.
\newblock \doi{10.1016/j.eml.2020.100839}.
\newblock URL \url{http://doi.org/10.1016/j.eml.2020.100839}.

\bibitem[Yang et~al.(2012)Yang, Kalnay, and Hunt]{Yang_2012}
S.-C. Yang, E.~Kalnay, and B.~Hunt.
\newblock Handling nonlinearity in an ensemble {K}alman filter: {E}xperiments
  with the tree-variable {L}orenz model.
\newblock \emph{Montly Weather Review}, 140\penalty0 (8):\penalty0 2628--2646,
  2012.

\bibitem[Yang and Church(2005)]{Yang_2005}
X.~Yang and C.~C. Church.
\newblock A model for the dynamics of gas bubbles in soft tissue.
\newblock \emph{The Journal of the Acoustical Society of America}, 118\penalty0
  (6):\penalty0 3595--3606, Dec 2005.
\newblock \doi{10.1121/1.2118307}.
\newblock URL \url{http://dx.doi.org/10.1121/1.2118307}.

\end{thebibliography}

\end{document}